\documentclass[10pt,notitlepage,onecolumn,showpacs,preprintnumbers,amsmath,amssymb,superscriptaddress, pra,floatfix]{revtex4-1}
\usepackage{bm}
\usepackage{graphicx}
\usepackage{dcolumn}
\usepackage{tabularx}
\usepackage{bm}
\usepackage{epstopdf}
\usepackage{amsmath}
\usepackage{color}
\usepackage{here}
\usepackage{setspace}
\usepackage[left=37.5truemm,right=37.5truemm]{geometry}

\begin{document}
\setstretch{1.0}
 
\preprint{Preprint}
\title{Emulating the local Kuramoto model with an injection-locked photonic crystal laser array}

\author{N. Takemura} 
\affiliation{Nanophotonics Center, NTT Corp., 3-1, Morinosato Wakamiya Atsugi, Kanagawa 243-0198, Japan}
\affiliation{NTT Basic Research Laboratories, NTT Corp., 3-1, Morinosato Wakamiya Atsugi, Kanagawa 243-0198, Japan}
\author{K. Takata}
\author{M. Takiguchi}
\author{M. Notomi}
\email[E-mail: ]{masaya.notomi.mn@hco.ntt.co.jp}
\affiliation{Nanophotonics Center, NTT Corp., 3-1, Morinosato Wakamiya Atsugi, Kanagawa 243-0198, Japan}
\affiliation{NTT Basic Research Laboratories, NTT Corp., 3-1, Morinosato Wakamiya Atsugi, Kanagawa 243-0198, Japan}

\begin{abstract}
The Kuramoto model is a mathematical model for describing the collective synchronization phenomena of coupled oscillators. We theoretically demonstrate that an array of coupled photonic crystal lasers emulates the Kuramoto model with non-delayed nearest-neighbor coupling (the local Kuramoto model). Our novel strategy employs indirect coupling between lasers via additional cold cavities. By installing cold cavities between laser cavities, we avoid the strong coupling of lasers and realize ideal mutual injection-locking with effective non-delayed dissipative coupling. First, after discussing the limit cycle interpretation of laser oscillation, we demonstrate the synchronization of two indirectly coupled lasers by numerically simulating coupled-mode equations. Second, by performing a phase reduction analysis, we show that laser dynamics in the proposed device can be mapped to the local Kuramoto model. Finally, we briefly demonstrate that a chain of indirectly coupled photonic crystal lasers actually emulates the one-dimensional local Kuramoto chain. We also argue that our proposed structure, which consists of periodically aligned cold cavities and laser cavities, will best be realized by using state-of-the-art buried multiple quantum well photonic crystals.
\end{abstract}
\maketitle

\section*{Introduction}
Nowadays, the investigation of synergetic dynamics emerging from coupled oscillators is an interdisciplinary study intensively discussed in physics, mathematics, chemistry, biology, and neuroscience \cite{Pikovsky2003}. Collective phenomena in coupled oscillators were investigated for the first time by Kuramoto, who used a large set of fully-connected oscillators which is a mathematical model called the Kuramoto model \cite{Kuramoto2003,Acebron2005}. In spite of the simplicity of the Kuramoto model, it comprises rich physics. For example, it exhibits a phase transition-like phenomenon from an incoherent state to a fully synchronized state when coupling strength reaches a threshold. Investigations of the Kuramoto model are not limited to theoretical ones. Actually, it is the simplest model for understanding various collective synchronization phenomena observed in nature, such as the collective synchronizations of neural oscillations and fireflies. Another actively studied direction is the emulation of the Kuramoto model with physical systems, for which the well-known example is the Josephson junction array \cite{Tsang1991,Wiesenfeld1996,Barbara1999}, though another promising approach is to use an array of coupled lasers \cite{Winful1988,Wang1988,Thornburg1997,Hohl1999,Kozyreff2000,Allaria2001,Rogister2004,Utsunomiya2011,Takata2012,Utsunomiya2015,Sun2019}. In this paper, we employ the latter approach and focus on a nanophotonic device, which provides an attractive playground for studying dynamical systems, with which synchronization of limit cycle oscillations has been theoretically and experimentally investigated \cite{Baas2008,Zhang2012,Bagheri2013,Walter2014,Ohadi2016,Xu2019}.

We propose a novel nanophotonic device that emulates the Kuramoto model with non-delayed nearest-neighbor coupling \cite{Sakaguchi1987,Daido1988,Strogatz1988,Hong2005,Acebron2005,Lee2010}, which we call the local Kuramoto model. Our idea is inspired by pioneering studies on coupled photonic crystal (PhC) lasers \cite{Altug2004,Altug2005,Hamel2015,Takata2017,Marconi2018,Takata2020} and by the mutual injection locking technique in laser physics \cite{Noda1990,Chan2003,TaukePedretti2011,Kurtz2005,Chen2008,Sun2015,Utsunomiya2015}. Different from the conventional injection-locking \cite{Kobayashi1981,Siegman1986}, mutual injection-locking involves neither master nor slave lasers. Our proposed device employs PhC lasers indirectly coupled via additional cold cavities. We demonstrate that the cold cavities play a crucial role in avoiding strong coupling between lasers, which results in ideal mutual injection-locking and dramatically simplifies the phase dynamics of laser oscillations. Compared with the other systems, nanophotonic Kuramoto models can be very compact devices that operate even at room temperatures. Furthermore, using PhC lasers, we aim for an on-chip realization of the local Kuramoto model, which may have an application as a coherent high-power laser. Additionally, in contrast to delayed coupling due to optical paths in free-space injection-locking \cite{Kozyreff2000,Utsunomiya2015}, our on-chip local Kuramoto model can provide stable coupling without coupling delay thanks to the direct evanescent coupling. Actually, the realization of dissipative coupling without time delay will be very difficult without using our scheme.

First, as a starting point, we consider two coupled PhC lasers coupled via a cold cavity. For this purpose, we interpret laser oscillation as limit cycle oscillation and model it by the Stuart-Landau equation. With coupled-mode equations, we numerically demonstrate the synchronization (mutual injection-locking) of two lasers. Furthermore, we confirm that strong-coupling between the two lasers is actually prohibited by the presence of the additional cold cavity. Second, in the same way as in our previous paper \cite{Takemura2020a}, we perform a phase reduction analysis to calculate  the phase equations of motion for two indirectly coupled lasers \cite{Kuramoto2003,Nakao2016,Stankovski2017}. The obtained phase equations of motion indicate that the phase dynamics of lasers indirectly coupled via cold cavities is equivalent to the local Kuramoto model. Finally, we demonstrate that a one-dimensional chain of indirectly coupled PhC lasers can emulate the one-dimensional local Kuramoto chain \cite{Zheng1998}. 

We also argue that our proposed device can be realized best by using buried multiple quantum well (MQW) PhC cavities \cite{Takeda2013,Matsuo2013,Takiguchi2016,Takemura2019}, where MQWs are locally embedded in a PhC slab. With this state-of-the-art technology, laser and cold cavities can be periodically aligned on a PhC chip.

\section*{Lasers as limit cycle oscillators}
Here, we review limit cycle interpretation for laser oscillation. In general, using complex field $\alpha$ and carrier number $N$, single-mode laser dynamics are, in the nonrotating frame, described by the following rate equations: \cite{Rice1994,Druten2000,Lariontsev2011}
\begin{eqnarray}
\dot{\alpha}&=&-i\omega_c\alpha-\frac{1}{2}\gamma_c\alpha+\frac{1}{2}\beta\gamma_\|N\alpha\label{eq:rate_alpha}\\
\dot{N}&=&-\gamma_\|N-\beta\gamma_\|N|\alpha|^2+P,\label{eq:rate_N}
\end{eqnarray}
where $P$ is the pumping rate for carriers, and $\omega_c$ is the resonance frequency of the laser cavity. Decay rates $\gamma_c$ and $\gamma_\|$ are photon and carrier decay rates, respectively. Note that, in this paper, by employing the quantum optics convention, the electric field rotates as $\alpha(t)=\alpha(0)e^{-i\omega_ct}$, which is opposite to the rotation in conventional coupled-mode equations, [$\alpha(t)=\alpha(0)e^{i\omega_ct}$]. The coefficient $\beta$ represents the fraction of photons spontaneously emitted into a lasing mode, and it is called the spontaneous emission coupling coefficient \cite{Rice1994}.
For simplicity, we neglect the linewidth enhancement factor in the rate equations (\ref{eq:rate_alpha}) and (\ref{eq:rate_N}) in the main text. In Section 5 in the supplemental material, we discuss the effect of the linewidth enhancement factor on synchronization, which may be negligible in quantum-dot lasers but generally has non-negligible effects in semiconductor lasers.
Here, it is worth noting that, in Eqs. (\ref{eq:rate_alpha}) and (\ref{eq:rate_N}), the terms $\frac{1}{2}\beta\gamma_\|N\alpha$ and $-\beta\gamma_\|N|\alpha|^2$ represent the stimulated emission, while there are no spontaneous emission terms. The effect of spontaneous emission will be included in the rate equations through a field noise term, if necessary. It is also important to note that Eqs. (\ref{eq:rate_alpha}) and (\ref{eq:rate_N}) hold only for a low $\beta(\ll1)$, which is usually the case in most lasers. The rate equations (\ref{eq:rate_alpha}) and (\ref{eq:rate_N}) are known to exhibit Hopf bifurcation, which is equivalent to lasing, when the pump rate reaches a lasing threshold $P=P_{\rm th}=\gamma_c/\beta$. 

\begin{figure}
\includegraphics[width=0.9\textwidth]{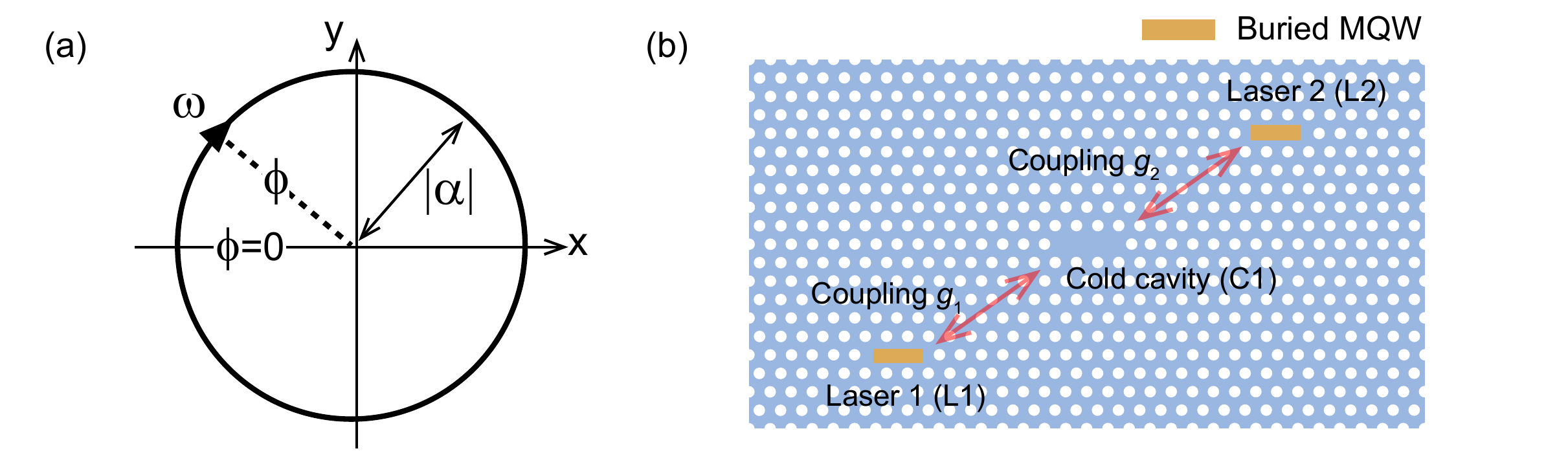}
\caption{(a) Laser oscillation is interpreted as limit cycle oscillation in the nonrotating frame, where the laser frequency $\omega$ corresponds to the frequency of the limit cycle. With the amplitude $|\alpha|=\sqrt{\varepsilon/\beta}$ and phase $\phi=\omega t$ of the laser, we define the limit cycle orbit as $(x(\phi),y((\phi)))=\sqrt{\varepsilon/\beta}(-\cos\phi,\sin\phi)$, where $x$ and $y$ are the real ${\rm Re}[\alpha]$ and imaginary parts ${\rm Im}[\alpha]$ of the field, respectively. (b) Illustration of two PhC lasers (L1 and L2) indirectly coupled via a cold cavity (C1). The three cavities are evanescently coupled with coupling strengths $g_1$ and $g_2$.}
\label{fig:limit_cycle}
\end{figure}

In this paper, for further simplification, we consider the case where the photon lifetime is much longer than the carrier lifetime ($\gamma_c\ll\gamma_\|$), which is called the class-A condition \cite{Arecchi2012}. With this assumption, we adiabatically eliminate the carrier degree of freedom as $\dot{N}=0$ \cite{Louisell1973,Haken1977}. The adiabatic elimination of the carrier dynamics reduces the rate equations (\ref{eq:rate_alpha}) and (\ref{eq:rate_N}) to
\begin{equation}
\dot{\alpha}=-i\omega_c\alpha+\frac{1}{2}\gamma_c\varepsilon\alpha-\frac{1}{2}\beta\gamma_c|\alpha|^2\alpha.\label{eq:SL}
\end{equation}
Equation (\ref{eq:SL}) is the well-known Stuart-Landau equation \cite{Kuramoto2003,Strogatz2018}, which is also called the Van der Pol equation \cite{Lee2013,Walter2014}. Importantly, parameter $\varepsilon$ in Eq. (\ref{eq:SL}) is the pump parameter defined as 
\begin{equation}
\varepsilon\equiv\frac{P-P_{\rm th}}{P_{\rm th}}\ \ {\rm with}\ \ P_{\rm th}=\frac{\gamma_c}{\beta}, 
\end{equation}
which indicates that the Hopf bifurcation (lasing) again occurs when $\varepsilon$ exceeds zero. Actually, when $\varepsilon>0$, the field amplitude $|\alpha|$ [see Fig. \ref{fig:limit_cycle}(a)] increases with an increase in the pump parameter as
\begin{equation}
|\alpha|=\sqrt{\frac{\varepsilon}{\beta}}\ \ {\rm for}\ \ \varepsilon\geq0. 
\end{equation}
Therefore, in Eq. (\ref{eq:SL}), the linear $\gamma_c\varepsilon\alpha/2$ and nonlinear term $\beta\gamma_c|\alpha|^2\alpha/2$ can be interpreted as gain and gain saturation, respectively. Here, it is important to stress that the laser oscillation itself is interpreted as limit cycle oscillation, and thus the resonance frequency of the laser cavity $\omega_c$ is the oscillation frequency of the limit cycle. As limit cycle oscillation emerges only in a nonlinear dissipative system with energy injection, lasing is achieved with the cavity decay, pumping, and gain saturation (nonlinearity).

Finally, we briefly comment on the effect of photon-carrier dynamics on synchronization properties, which will be important in real PhC cavity lasers. Since PhC cavity lasers are semiconductor lasers, their carrier lifetime is much longer than the photon lifetime (sometimes called class-B lasers \cite{Arecchi2012}), and the relaxation oscillation appears around lasing threshold \cite{Takemura2012,Wang2015}. Therefore, in real PhC cavity lasers, the adiabatic elimination approximation of the carrier degree of freedom cannot be justified, and we need to directly simulate the rate equations (\ref{eq:rate_alpha}) and (\ref{eq:rate_N}). Fortunately, we found that Eqs. (\ref{eq:rate_alpha}) and (\ref{eq:rate_N}) quantitatively provide the same results as the Stuart-Landau equation as long as phase dynamics are concerned, which can also be confirmed with the phase equation of motion for the class-B rate equations. See Section 4 in the supplemental material.

\section*{Synchronization of two lasers}
\subsection*{Coupled-mode equations}
Now, we consider the device shown in Fig. \ref{fig:limit_cycle}(b), where the two lasers (L1 and L2) are indirectly coupled via the cold cavity (C1). The corresponding coupled-mode equations of motion representing field dynamics are given by 
\begin{eqnarray}
\dot{\alpha}_1&=&-i\omega_1\alpha_1+\frac{1}{2}\gamma_1\varepsilon_1\alpha_1-\frac{1}{2}\beta_1\gamma_1|\alpha_1|^2\alpha_1-ig_1E_1\label{eq:two_1}\\
\dot{E}_1&=&-i\Omega_1E_1-\frac{1}{2}\Gamma_1E_1-ig_1\alpha_1-ig_2\alpha_2\label{eq:two_2}\\
\dot{\alpha}_2&=&-i\omega_2\alpha_2+\frac{1}{2}\gamma_2\varepsilon_2\alpha_2-\frac{1}{2}\beta_2\gamma_2|\alpha_2|^2\alpha_2-ig_2E_1,\label{eq:two_3}
\end{eqnarray}
where $\alpha_{1,2}$ and $E_1$ represent fields in the laser cavity and cold cavity, respectively. Additionally, $\omega_{1,2}$ and $\Omega_1$ respectively represent the resonance frequencies of the laser cavities  (L1 and L2) and coldcavity (C1). Similarly, $\gamma_{1,2}$ and $\Gamma_1$ are the field decay rates of the laser- (L1,2) and cold  cavity (C1), respectively. The parameter $\beta_{1,2}$ is the spontaneous emission coupling coefficient, while $\varepsilon_{1,2}$ is the pump parameter for laser L1 and L2. Finally, the two coupling strengths between the cavities are denoted by $g_1$ and $g_2$. For simplicity, in the rest of this paper, we use $\beta_1=\beta_2=0.001$ and $\varepsilon_1=\varepsilon_2=1.0$, which is above the lasing threshold. Furthermore, we use the same values for the normalized decay rates of the laser cavity and cold cavity: $\Gamma_1=\gamma_2=\gamma_1\equiv1$, where $\gamma_1$ is interpreted as a dimensionless parameter for numerical simulations.

To observe synchronization, we set the resonance frequencies of the two laser cavities as $\omega_2=\omega_1+\Delta\omega$ with $\Delta\omega=0.01\gamma_1$, where $\Delta\omega\equiv\omega_2-\omega_1$ is the frequency difference between the two lasers. For the cold cavity (C1), for simplicity, we use the same resonance frequency as L1: $\Omega_1=\omega_1$.

\subsection*{Time evolutions}
By showing field time evolutions described by the coupled-mode equations (\ref{eq:two_1})-(\ref{eq:two_3}), we demonstrate the synchronization of two lasers (mutual injection locking). Since the typical laser frequency, which is on the order of hundreds of terahertz, we perform the rotating-frame transformation for all fields, for example, as $\alpha_1e^{-i\omega_s t}\rightarrow\alpha_1$. With this rotating frame transformation, we shift the resonance frequencies of the cavities as $\omega_1^\prime\equiv\omega_1-\omega_s=1\gamma_1$, $\omega_2^\prime\equiv\omega_2-\omega_s=1.01\gamma_1$, and $\Omega_1^\prime\equiv\Omega_1-\omega_s=1\gamma_1$. Importantly, there is an arbitrariness in the absolute frequencies, and only the relative frequencies are important. Thus, the frequency of the rotating frame, $\omega_s$, is arbitrary, and only the relative values between $\omega_1$, $\omega_2$, and $\Omega_1$ matter. 
\begin{figure}
\includegraphics[width=0.9\textwidth]{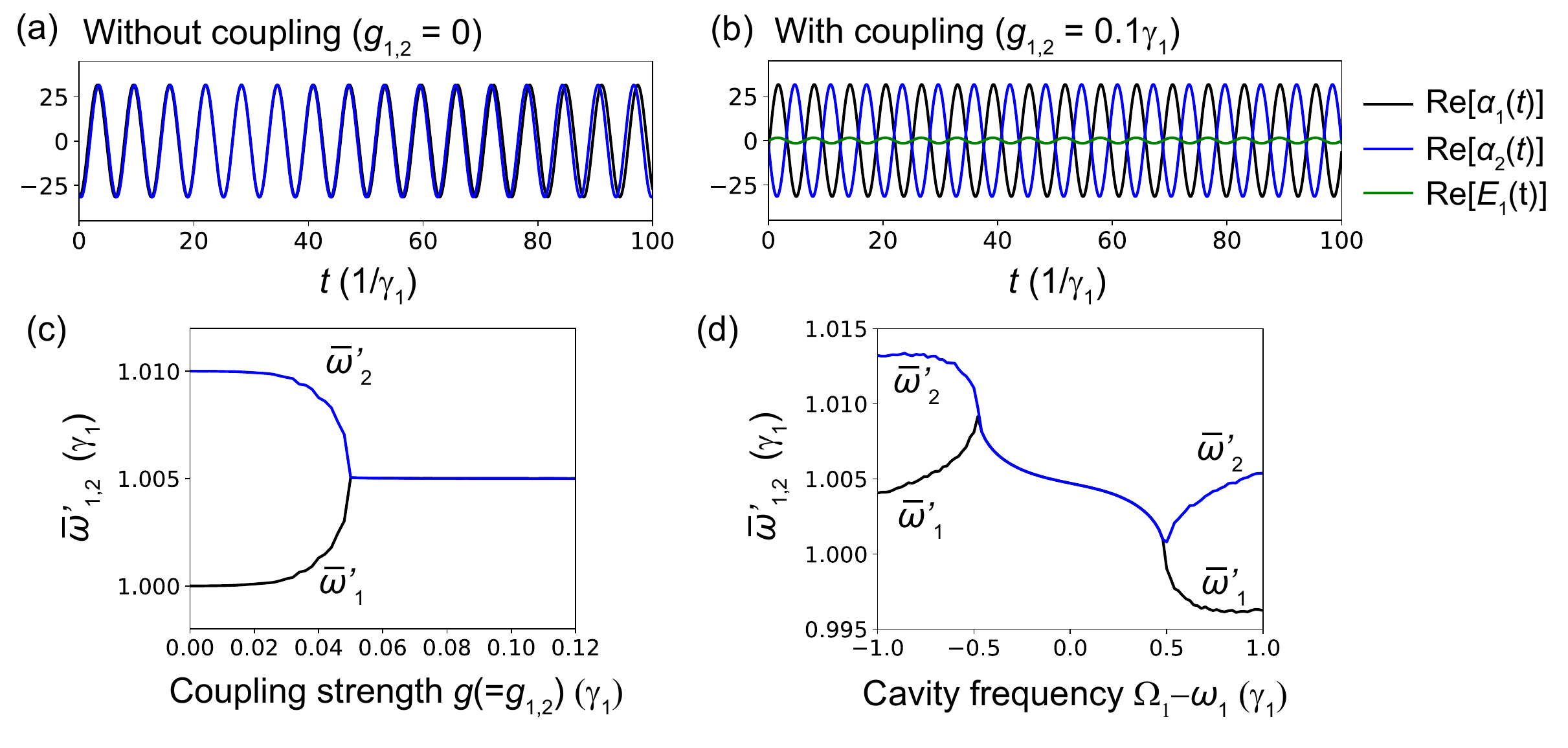}
\caption{Simulations for two lasers [see Fig. \ref{fig:limit_cycle}(b)] coupled via a cold cavity. The simulated time evolutions of the real part of the field ${\rm Re}[\alpha_{1,2}(t)]$ without $g_{1,2}=0$ (a) and with coupling $g_{1,2}=0.1\gamma_1$ (b). Here, we used the shifted laser and cold cavity frequencies $\omega_1^\prime=1\gamma_1$, $\omega_2^\prime=1.01\gamma_1$, $\Omega_1^\prime=1\gamma_1$. In (b), we also show the time evolution of the real part of the field of the cold cavity ${\rm Re}[E_{1}(t)]$. (c) Mean frequency of the laser oscillation $\bar{\omega}_{1,2}$ as a function of coupled strength $g_{1,2}$. (d) Mean frequency $\bar{\omega}_{1,2}$ for fixed coupling strength ($g_{1,2}=0.1\gamma_1$) but as a function of the resonance frequency of the cold cavity $\Omega_1$, which tunes the effective coupling strength between the two lasers.}
\label{fig:sync_lasers}
\end{figure}

First, Fig. \ref{fig:sync_lasers}(a) shows the time evolutions of the real parts of the fields ${\rm Re}[\alpha_1(t)]$ (black) and ${\rm Re}[\alpha_2(t)]$ (blue) for the lasers L1 and L2, respectively, without coupling between cavities $g_1=g_2=0$. Without coupling between the cavities, there is no photon in cold-cavity C3, and thus $E_1(t)=0$. As we expect, in Fig. \ref{fig:sync_lasers}(a), the fields in laser L1 and L2 oscillate with their own frequencies: $\omega_1^\prime=1\gamma_1$ and $\omega_2^\prime=1.01\gamma_1$. Second, we introduce coupling between the cavities as $g_1=g_2=0.1\gamma_1$ in Fig. \ref{fig:sync_lasers}(b), where the green curve is the time evolution of the real part of the field in cold-cavity C1 ${\rm Re}[E_1]$. Figure \ref{fig:sync_lasers}(b) indicates that the two indirectly coupled laser oscillations exhibit synchronization (mutual injection locking), which is the main result of this paper. Furthermore, the synchronization phase is anti-phase, which is called anti-phase synchronization. Importantly, thanks to cold-cavity C3, normal-mode splitting associated with strong coupling between the two lasers is prohibited, which is confirmed from the fact that the frequency of the synchronized oscillations does not depend on the initial states of two lasers (not shown). In fact, when the system is in the strong-coupling regime, depending on the initial states of two lasers, for example, they form a ``bonding" or ``anti-bonding" mode, and their frequencies become lower or higher than the original oscillation frequencies \cite{Takemura2020a}. Importantly, no matter how weak the coupling is, directly coupled lasers  exhibit normal-mode splitting because they have no decay (gain). Note that, for Eqs. (\ref{eq:two_1})-(\ref{eq:two_3}), anti-phase synchronization always occurs for any initial state, while if the signs of the two couplings are opposite such as $g_2=-g_1$, in-phase synchronization always occurs (not shown) The sign of a coupling constant depends on the overlap integral of cavity fields and may vary depending on the distance between cavities. In the device design in this paper, since all the distances between cavities are designed to be equal, all the signs of coupling constants can be assumed to be the same. In any case, the property that a synchronization phase does not depend on initial phases of lasers is of importance because the initial phase of PhC lasers cannot be controlled experimentally.

\subsection*{Synchronization tree}
In Fig. \ref{fig:sync_lasers}(c), we show the mean frequencies of the two laser oscillations $\bar{\omega}_{1}^\prime$ and $\bar{\omega}_{2}^\prime$ as a function of the coupling between cavities $g_{1,2}$. Since, in general, limit cycle oscillations are quasi-periodic when coupling strength is lower than the critical strength of synchronization, we need to use their mean frequencies obtained with peak detection. Figure \ref{fig:sync_lasers}(c) clearly indicates that the mean frequencies symmetrically approach each other with an increase in the coupling strength $g$ and that they merge as $\bar{\omega}_1^\prime=\bar{\omega}_2^\prime=1.005\gamma_1$ at the critical strength $g=0.05\gamma_1$. In fact, the frequency $\bar{\omega}_{1,2}^\prime=1.005\gamma_1$ is the mean frequency of $\omega_1^\prime=1\gamma_1$ and $\omega_2^\prime=1.01\gamma_1$ without coupling. Note that the synchronization tree shown in Fig. \ref{fig:sync_lasers}(c) is approximately symmetric for $\bar{\omega}_{1}^\prime$ and $\bar{\omega}_{2}^\prime$, which is because the parameters are almost the same for L1 and L2. 

Furthermore, in Fig. \ref{fig:sync_lasers}(d), we plot the mean frequency $\bar{\omega}_{1,2}^\prime$ as a function of the resonance frequency of the cold cavity $\Omega_1$, where the coupling strengths are fixed as $g_1=g_2=0.07\gamma_1$ while $\Omega_1$ is swept from $\omega_1-1\gamma_1$ to $\omega_1+1\gamma_1$. Figure \ref{fig:sync_lasers}(d) indicates that the effective coupling strengths between the cavities can be tuned by changing the resonance frequency of the cold cavity $\Omega_1$. Intuitively, as the cold-cavity's frequency deviates from the resonance frequencies of the two lasers, the effective coupling strengths decrease. In PhC cavities, the tuning of cavity coupling strength, which is determined by the distance between cavities, is almost impossible. Meanwhile the tuning of the cold cavity's resonance frequency is technically available with the carrier-injection \cite{Fushman2007,Tanabe2009} or thermo-optic techniques \cite{Chong2004,Faraon2009}, and thus the synchronization tree shown in Fig. \ref{fig:sync_lasers}(d) could be measured. 

\section*{Phase equations of motion}
In this section, as we did in Ref. \cite{Takemura2020a}, by performing the phase reduction analysis \cite{Winfree1967,Kuramoto2003} for Eqs. (\ref{eq:two_1})-(\ref{eq:two_3}), we attempt to obtain phase equations of motion. In our case, the phase of limit cycle oscillation is nothing else but the phase of a laser $\phi$ as illustrated in Fig. \ref{fig:limit_cycle}(a), and thus the interpretation of corresponding phase equations of motion is also straightforward. Furthermore, we show that the determination of phase equations of motion is of importance in terms of mapping our model to the local Kuramoto model. The price to pay for obtaining phase equations of motion is the adiabatic elimination of the field in cold-cavity C1, which is required to transform the indirectly coupled system to a directly coupled model with dissipative coupling. 

\subsection*{Adiabatic elimination approximation}
The adiabatic elimination of the cold-cavity field degree of freedom $\dot{E}_1=0$ requires that field $E_1$ rapidly decays compared with the laser field $\alpha_{1,2}$, and thus $E_1$ adiabatically follows $\alpha_{1}$ and $\alpha_{2}$. The time-scale of a variable is generally characterized by its decay rate. Therefore, the conventional adiabatic elimination of field $E_1$ requires that the decay rate $\Gamma_1$ must be larger than the decay rates of $\alpha_1$ and $\alpha_2$, as shown in Section. 2, which is not the case, for example, when we consider $\Gamma_1=\gamma_{1,2}$ as in Fig. \ref{fig:sync_lasers}. However, importantly, the time scale of the laser field $\alpha_{1,2}$ is not characterized solely by $\gamma_{1,2}$. Now, it is important to define the effective decay rates for $\alpha_1$, $\alpha_2$, and $E_1$, including both oscillation frequencies and pump parameters, as $\lambda_1\equiv-\gamma_1\epsilon_1/2+i\omega_1$, $\lambda_2\equiv-\gamma_2\epsilon_2/2+i\omega_1$, and $\Lambda_1\equiv\Gamma_1/2+i\Omega_1$, respectively. Here, the oscillation frequencies $\omega_1$, $\omega_2$, and $\Omega_1$ are the imaginary parts of the effective decay rates. First, as pointed out by Haken \cite{Haken1977}, to compare the time scales of the variables, in the effective decay rates, the imaginary parts must be negligible compared to the real parts: ${\rm Im}[\lambda_{1,2}]\ll{\rm Re}[\lambda_{1,2}]$ and ${\rm Im}[\Lambda_{1}]\ll{\rm Re}[\Lambda_{1}]$. Even though the cavity resonance frequencies $\omega_{1,2}$ and $\Omega_1$ are always much higher than the terms $\gamma_{1,2}\epsilon_{1,2}/2$ and $\Gamma_{1}/2$, if all the resonance frequencies of the cavities have similar values $\omega_1\simeq\omega_2\simeq\Omega_1$, the imaginary parts in the effective decay rates become negligible in a rotating frame with the frequency of $\Omega_1$. Second, by comparing the real parts of the effective decay rates ${\rm Re}[\lambda_{1,2}]$ and ${\rm Re}[\Lambda_{1}]$, we find that the sign of ${\rm Re}[\Lambda_{1}]$ is always positive, while the sign of ${\rm Re}[\lambda_{1,2}]$ can be negative due to gain when the pump power is above the threshold $\varepsilon_{1,2}\geq0$. According to Ref. \cite{Haken1977,Haken1993}, when ${\rm Re}[\Lambda_{1}]>0$ and ${\rm Re}[\lambda_{1,2}]\leq0$, the field $E_1$ is a ``stable" mode that rapidly decays, while the laser fields $\alpha_1$ and $\alpha_2$ are unstable modes that do not decay but govern the slow dynamics of the system, which allows putting $\dot{E}_1=0$ (adiabatic elimination). In fact, the unstable mode $\alpha_{1,2}$ ``enslaves" the stable mode $E_1$ and plays a role as an ``order parameter" (the slaving principle \cite{Haken1977,Haken1993}). 

\begin{figure}
\includegraphics[width=0.9\textwidth]{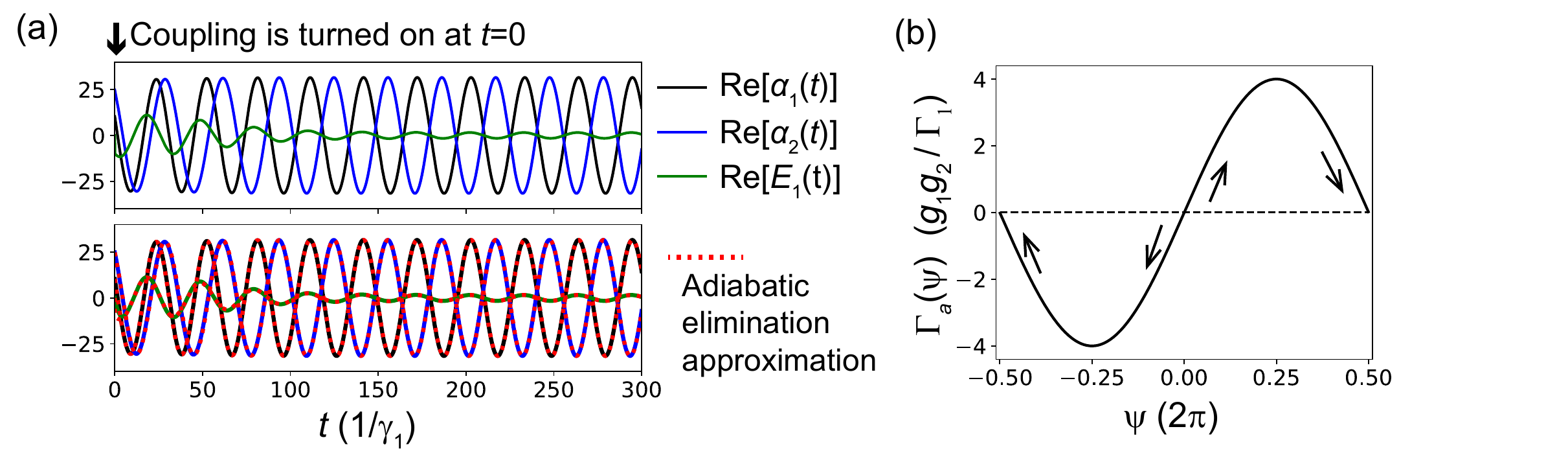}
\caption{(a) Time evolution of the fields ${\rm Re}[\alpha_{1,2}(t)]$ and ${\rm Re}[E(t)_{1}]$, but coupling ($g_{1,2}=0.1\gamma_1$) is turned on at $t=0$, which represents synchronization dynamics. The upper panel shows the time evolutions of the fields calculated with the original coupled-mode equations. In the lower panel, the time evolutions of the fields calculated with the adiabatic elimination approximation are shown as red dashed lines with the original plots. The shifted frequencies of the laser and cold cavities are $\omega_1^\prime=0.2\gamma_1$, $\omega_2^\prime=0.21\gamma_1$, and $\Omega_1^\prime=0.2\gamma_1$. (b) Anti-symmetric part of the phase coupling function $\Gamma_a(\psi)$ given by Eq. (\ref{eq:phase_eq_final}), where $\psi$ is the phase difference between the two laser phases defined as $\psi\equiv\phi_2-\phi_1$.}
\label{fig:gamma}
\end{figure}

Now, setting $\dot{E}_1=0$ for Eq. (\ref{eq:two_2}), we eliminate the cold-cavity field degree of freedom as
\begin{eqnarray}
E_1=-i\frac{2}{\Gamma_1}(g_1\alpha_1+g_2\alpha_2).
\label{eq:eliminated}
\end{eqnarray}
By substituting Eq. (\ref{eq:eliminated}) into Eqs. (\ref{eq:two_1}) and (\ref{eq:two_3}), we obtain approximated equations of motion: 
\begin{eqnarray}
\dot{\alpha}_1&=&-i\omega_1\alpha_1+\left[\frac{1}{2}\gamma_1\varepsilon_1-\frac{2g_1^2}{\Gamma_1}\right]\alpha_1-\frac{1}{2}\beta_1\gamma_1|\alpha_1|^2\alpha_1-\frac{2g_1g_2}{\Gamma_1}\alpha_2\label{eq:ad_1}\\
\dot{\alpha}_2&=&-i\omega_2\alpha_2+\left[\frac{1}{2}\gamma_2\varepsilon_2-\frac{2g_2^2}{\Gamma_1}\right]\alpha_2-\frac{1}{2}\beta_2\gamma_2|\alpha_2|^2\alpha_2-\frac{2g_1g_2}{\Gamma_1}\alpha_1.\label{eq:ad_2}
\end{eqnarray}
To confirm the validity of this adiabatic elimination approximation, in Fig. \ref{fig:gamma}(a), we show synchronization dynamics calculated both with the original equations of motion (\ref{eq:two_1})-(\ref{eq:two_3}) and approximated equations (\ref{eq:ad_1}) and (\ref{eq:ad_2}). In Fig. \ref{fig:gamma}(a), coupling with $g_{1,2}=0.1\gamma_1$ is switched on at $t=0$ for uncoupled steady-state laser oscillations, and thus the time evolutions of fields represent synchronization dynamics from the unsynchronized to synchronized state. The upper panel in Fig. \ref{fig:gamma}(a) shows only the synchronization dynamics calculated with the original equations of motion (\ref{eq:two_1})-(\ref{eq:two_3}). Meanwhile, in the lower panel, synchronizations calculated with the original equations of motion (solid lines) overlap those calculated with the approximated equations of motion (dashed lines), which clearly indicates that two time evolutions are almost indistinguishable and that the adiabatic elimination approximation is surprisingly good. Note that, to clearly show the synchronization dynamics in Fig. \ref{fig:gamma}(a), we used shifted frequencies $\omega_1^\prime=0.2\gamma_1$, $\omega_2^\prime=\omega_1^\prime+\Delta\omega=0.21\gamma_1$ and $\Omega_1^\prime=\omega_1^\prime=0.2\gamma_1$, which are lower than those Fig. \ref{fig:limit_cycle}. As we commented in Section 3.1, these shifts of the resonance frequencies do not change the physics, because only the relative relationship between the resonance frequencies is important. Since the field in the cold cavity was adiabatically eliminated, Eqs. (\ref{eq:ad_1}) and (\ref{eq:ad_2}) represent directly coupled lasers. Furthermore, in Eqs. (\ref{eq:ad_1}) and (\ref{eq:ad_2}), the effective couplings represented by $-(2g_1g_2/\Gamma_1)\alpha_2$ and $-(2g_1g_2/\Gamma_1)\alpha_1$ are non-energy-conserving dissipative couplings, which intuitively explains why normal-mode splitting does not appear in our model. Additionally, in Eqs. (\ref{eq:ad_1}) and (\ref{eq:ad_2}), the effective dissipative coupling does not have the time delay.  
 
Finally, we comment on synchronization with a large coupling strength. We found that Eqs. (\ref{eq:ad_1}) and (\ref{eq:ad_2}) fail to reproduce synchronization dynamics when $g_{1,2}\ge\gamma_{1,2},\Gamma_1$, which is because the adiabatic elimination approximation cannot describe coherent intensity oscillation between cavities associated with this parameter region [see Section 2 in the Supplemental Material (SM)]. Therefore, the complete conditions required for the adiabatic elimination approximation are
\begin{equation}
\omega_1\simeq\omega_2\simeq\Omega_1\ {\rm and}\ g_{1,2}<\gamma_{1,2},\Gamma_1.
\label{eq:ad_conditions}
\end{equation}
Here, it is also important to stress that, although the adiabatic elimination fails to describe synchronization dynamics, even when $g_{1,2}\ge\gamma_{1,2},\Gamma_1$, stable synchronization itself can occur and the adiabatic elimination approximation well reproduces the steady-state synchronized oscillations (see Section 2 in the SM). Furthermore, even when the coupling is extremely strong, for example, $g_{1,2}=10\gamma_1$, we can observe stable synchronization, where no normal-mode splitting is present (not shown). This insensitivity to coupling strength will be  advantageous in terms of real device designs, because adjusting the value of weak coupling strength is technically difficult \cite{Takemura2020a}. Furthermore, if coupling  is sufficiently strong, we may prove synchronization from spectral shapes, which is discussed again in Section 6. 

\subsection*{Phase reduction analysis}
Now, we perform the phase reduction analysis for equations of motion (\ref{eq:ad_1}) and (\ref{eq:ad_2}), which were obtained with the adiabatic elimination approximation. Here, we make use of the consequence of the phase reduction theory without going into the theoretical detail, which is briefly provided in Section 1 in the SM (further details can be found in our recent paper \cite{Takemura2020a} and in Refs \cite{Kuramoto2003,Nakao2016}). The objective of the phase reduction analysis is to obtain the phase equations of motion for the phases of the laser L1 ($\phi_{1}$) and L2 ($\phi_{2}$) represented as 
\begin{eqnarray}
\dot{\phi_1}&=&-\omega_1+\Gamma_{12}(\phi_1-\phi_2)\label{eq:phi1}\\
\dot{\phi_2}&=&-\omega_2+\Gamma_{21}(\phi_2-\phi_1),\label{eq:phi2}
\end{eqnarray}
where $\Gamma_{12}(\phi)$ and $\Gamma_{21}(\phi)$ are called the phase-coupling functions. For the approximated equations of motion (\ref{eq:ad_1}) and (\ref{eq:ad_2}), we found that $\Gamma_{12}(\phi)$ and $\Gamma_{21}(\phi)$ can be analytically calculated as
\begin{eqnarray}
\Gamma_{12}(\theta)=\Gamma_{21}(\theta)=\frac{2g_1g_2}{\Gamma_1}\sin\theta
\label{eq:Gamma_sin}
\end{eqnarray}

Finally, the phase difference between the two lasers $\psi\equiv\phi_2-\phi_1$ follows the following simple equation of motion:
\begin{equation}
\dot{\psi}=-\Delta\omega+\Gamma_a(\psi)\ \ {\rm with}\ \ \Gamma_a(\psi)=\frac{4g_1g_2}{\Gamma_1}\sin\psi,\label{eq:phase_eq_final}
\end{equation}
where $\Delta\omega\equiv\omega_2-\omega_1$ is the frequency difference between the two lasers already defined in Section 3.1. Here, $\Gamma_a(\psi)\equiv\Gamma_{21}(\psi)-\Gamma_{12}(-\psi)$ is the anti-symmetric part of the phase coupling function $\Gamma_{21}(\psi)$, which is shown in Fig. \ref{fig:gamma}(b). For a negligible laser frequency difference $\Delta\omega\simeq0$, since $\Gamma_a(\pi)=0$ and $\Gamma^\prime_a(\pi)<0$ hold for Eq. (\ref{eq:phase_eq_final}), phase locking occurs at the phase $\psi=\phi_2-\phi_1=\pi$, which is anti-phase synchronization as expected from the simulations [see the arrows in Fig. \ref{fig:gamma}(b)]. Meanwhile, since $\Gamma_a(0)=0$ and $\Gamma^\prime_a(0)>0$ hold for $\phi=0$, the phase $\phi=0$ is an unstable fixed point. Of course, for a non-negligible frequency difference $\Delta\omega\neq0$, the synchronization phase shifts from $\pi$. The phase equations of motion predict not only the synchronization phase but also the critical coupling strength of synchronization. For Eq. (\ref{eq:phase_eq_final}) to have a phase-locking solution, the condition $-4g_1g_2/\Gamma_1\leq\Delta\omega\leq4g_1g_2/\Gamma_1$ must be satisfied. For the oscillation frequency difference $\Delta\omega=0.01\gamma_1$ and cold-cavity decay rate $\Gamma_1=1\gamma_1$, which are assumed in Fig. \ref{fig:sync_lasers}(c), synchronization occurs when the coupling strengths reach $g_1=g_2=0.05\gamma_1$ [see Fig. \ref{fig:sync_lasers}(c)] because the above phase-locking condition is satisfied with these parameters as $4g_1g_2/\Gamma_1=0.01\gamma_1=\Delta\omega$. 
 
Furthermore, the analytically calculated phase coupling functions in Eq. (\ref{eq:Gamma_sin}) are also of importance for mapping our model to the local Kuramoto model. In fact, the phase equations of motion are explicitly written as 
\begin{eqnarray}
\dot{\phi_1}&=&-\omega_1+\tilde{g}_{12}\sin(\phi_1-\phi_2)\label{eq:Kuramoto1}\\
\dot{\phi_2}&=&-\omega_2+\tilde{g}_{21}\sin(\phi_2-\phi_1),\label{eq:Kuramoto2}
\end{eqnarray}
where $\tilde{g}_{ij}\equiv2g_ig_j/\Gamma_1$ ($\tilde{g}_{ij}=\tilde{g}_{ji}$) is the effective coupling strength. The phase equations of motion (\ref{eq:Kuramoto1}) and (\ref{eq:Kuramoto2}) are straightforwardly extended to a one-dimensional chain or two-dimensional array as
\begin{equation}
\dot{\phi_i}=-\omega_i+\sum_{j\in N_i}\tilde{g}_{ij}\sin(\phi_i-\phi_j),\label{eq:Kuramoto_array}
\end{equation}
where $N_j$ represents the nearest neighbour sites of the $i$th site. Importantly, the coupled phase oscillator described by Eq. (\ref{eq:Kuramoto_array}) is equivalent to the local Kuramoto model \cite{Daido1988,Acebron2005}. Note that, in the original Kuramoto model, the sign of the coupling is minus as $-\tilde{g}_{21}$, and thus in-phase synchronization occurs.

In conclusion, with the aide of the phase reduction theory, we proved that an array of lasers with cold-cavity-mediated coupling can emulate the nearest-neighbor coupled Kuramoto model (the local Kuramoto model). Note that, of course, the strict mapping of given coupled-mode equations to the local Kuramoto model (\ref{eq:Kuramoto_array}) requires an adiabatic elimination condition similar to Eq. (\ref{eq:ad_conditions}).

\section*{Array configuration}
\begin{figure}
\includegraphics[width=0.9\textwidth]{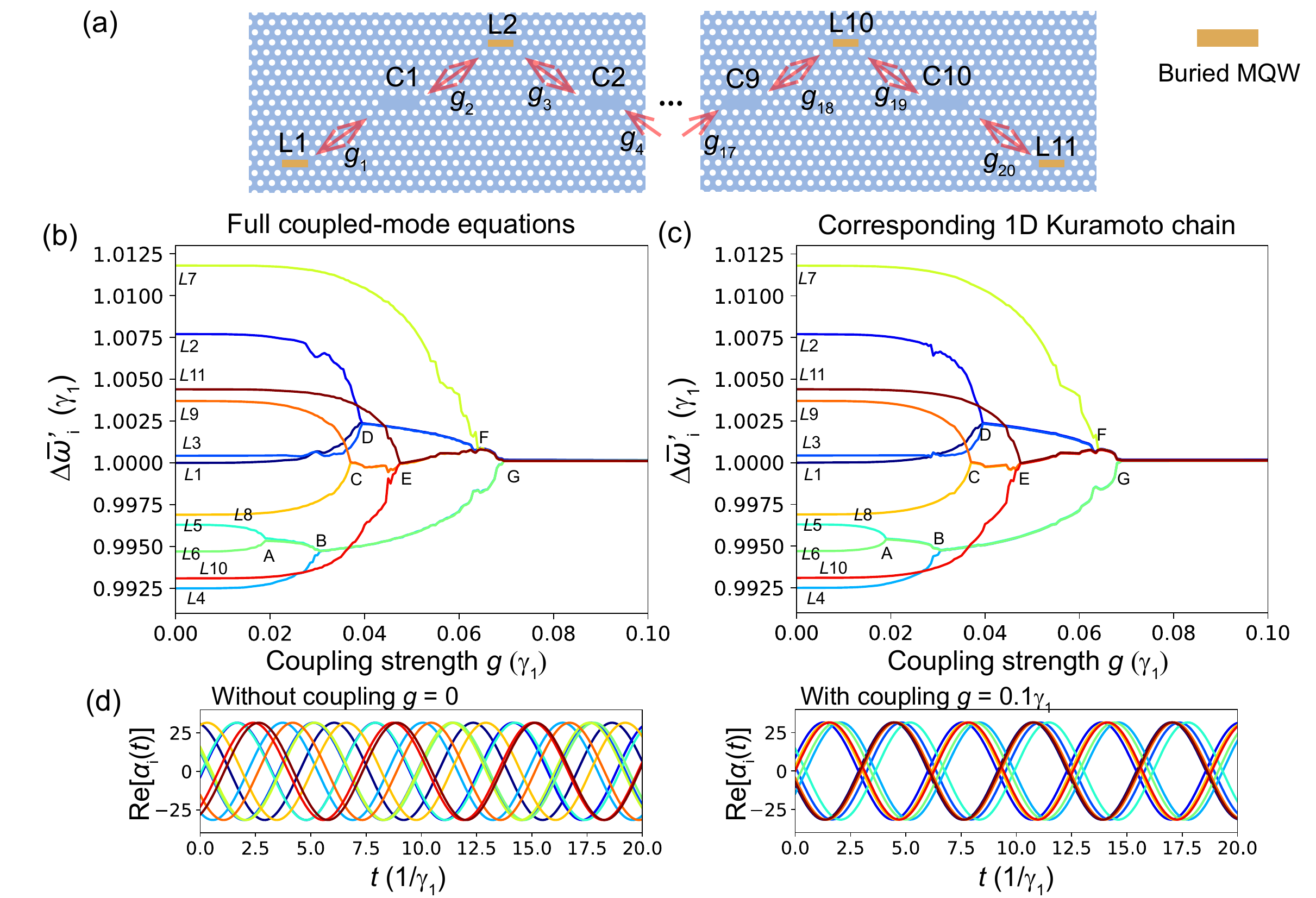}
\caption{(a) Schematic of a chain of eleven indirectly coupled PhC lasers that emulates the Kuramoto chain. Indices L$i$ and C$i$ represent the $i$th laser and cold cavities, respectively. The shifted resonance frequencies of the laser cavities are $\omega_1^\prime=$1.0000, $\omega_2^\prime=$1.0077, $\omega_3^\prime=$1.0004, $\omega_4^\prime=$0.9925, $\omega_5^\prime=$0.9963, $\omega_6^\prime=$0.9947, $\omega_7^\prime=$1.0118, $\omega_8^\prime=$0.9969, $\omega_9^\prime=$1.0037, $\omega_{10}^\prime=$0.9931, and $\omega_{11}^\prime=$1.0044, where the units are $\gamma_1$.
For the other parameters, we use $\beta_i=0.001$, $\varepsilon_i=1.0$, $\gamma_i\equiv1$, $\Omega_i=1\gamma_1$ and $\Gamma_i=\gamma_1\equiv1$ for all $i$.
(b) The mean oscillation frequencies of the eleven lasers $\bar{\omega}_i^\prime$ are shown as a function the coupling strength $g_i=g$ for all $i$. The synchronization points are denoted by A-G. (d) Time evolutions of the real parts of the fields in all the laser cavities without $g=0$ (left) and with coupling $g=0.1\gamma_1$ (right).}
\label{fig:tree}
\end{figure}

Although the investigation of rich physics emerging from coupled phase oscillators is beyond the scope of this paper, we briefly simulate a one-dimensional chain of indirectly coupled PhC lasers and demonstrate that our device can actually reproduce collective dynamics predicted for the one-dimensional local Kuramoto chain \cite{Zheng1998}. The chain of indirectly coupled PhC lasers is schematically illustrated in Fig. \ref{fig:tree}(a), where eleven laser cavities and ten cold cavities are alternately aligned. Of course, the configuration of cavities to realize the local Kuramoto chain is not limited to that shown in Fig. \ref{fig:tree}(a), and various configurations can be imagined. For a one-dimensional chain, in principle, even the periodic boundary condition may be implemented with a ring-like configuration. For simplicity, for all the laser cavities, we assume $\beta_i=0.001$, $\varepsilon_i=1.0$, and $\gamma_i\equiv1$. Similarly, all the cold cavities have the same resonance frequencies and photon decay rate: $\Omega_i=1\gamma_1$ and $\Gamma_i=\gamma_1\equiv1$ for all $i$. In Section 6 in the supplemental material, we demonstrate large-scale synchronization when the parameter values of all laser and cold cavities are slightly different. Furthermore, as in Section 3, we assume that all the coupling constants have the same strengths: $g_{i}=g$ for all $i$. Meanwhile, the resonance frequencies of the eleven laser cavities are randomly distributed around a mean frequency $\bar{\omega}_i=1\gamma_1$ [for the actual values of $\omega_i$, please see the caption of Fig. \ref{fig:tree}]. Note that since all the laser and cold cavities have similar resonance frequencies and the coupling strengths are smaller than the cavity decay rates, an adiabatic elimination condition similar to Eq. (\ref{eq:ad_conditions}) is satisfied, and thus corresponding simple phase equations of motion are expected to exist. 

By directly simulating the full coupled-mode equations corresponding to the configuration shown in Fig. \ref{fig:tree}(a), we calculated the mean frequencies of the laser oscillations as a function of the coupling strength $g$ [see the synchronization tree in Fig. \ref{fig:tree}(b)]. As Fig. \ref{fig:tree}(b) indicates, with an increase in coupling strength $g$, synchronized clusters are gradually formed, and finally all clusters merge into a single fully synchronized cluster at $g\simeq0.07\gamma_1$ [see G on Fig. \ref{fig:tree}(b)]. Similarly to Ref. \cite{Zheng1998}, when two [at A, B, C in Fig. \ref{fig:tree}(b)] or three [at D, E in Fig. \ref{fig:tree}(b)]  adjacent oscillators (or clusters) have close oscillation frequencies, they form a new synchronized cluster with an increase in coupling strength. When adjacent clusters have largely different frequencies, while non-adjacent clusters have similar frequencies, the non-adjacent clusters form a synchronized cluster. In fact, the synchronization denoted by F in Fig. \ref{fig:tree}(b) consists of the non-adjacent oscillators (clusters) L1-3 and L7-11. Furthermore, in Fig. \ref{fig:tree}(c), we show a synchronization tree calculated with the local Kuramoto chain [Eq. (\ref{eq:Kuramoto_array})] corresponding to Fig. \ref{fig:tree}(b). The fact that both synchronization trees have almost the same structures indicates that our proposed device will actually emulate the local Kuramoto model. 

Finally, the time evolutions of the laser oscillations without ($g=0$) and with coupling ($g=0.1\gamma_1$) are shown in the left and right panels of Fig. \ref{fig:tree}(d), respectively. When there is no coupling, as we expect, the laser oscillations are totally uncorrelated, while all the laser oscillations are fully synchronized with coupling $g=0.1\gamma_1$. Interestingly, in this fully synchronized state [see the right panel in Fig. \ref{fig:tree}(d)], the phases are opposite between the even and odd sites of the lasers oscillations. Therefore, even in the one-dimensional chain, a pair of adjacent laser oscillations exhibit anti-phase synchronization. Note that, in Fig. \ref{fig:tree}(c) and (d), the ``de-synchronization" discovered in \cite{Zheng1998} was not observed, which may be due to the small number of oscillators or, more interestingly, could be associated with anti-phase synchronization. We also comment on the offsets of the synchronization phases in the fully-synchronized oscillations shown in the right panel of Fig. \ref{fig:tree}(d), where the synchronization phases slightly differ depending on the pair of the synchronized oscillations. We found that, with a further increase in coupling strength, these offsets of the synchronization phases disappear and that all the pairs of synchronized oscillations become indistinguishable.

\section*{Discussion}
Here, we discuss several details that will be of importance in a real device design and experiments. In experiments, the easiest method to observe synchronization may be the spectral measurement of laser emissions. Since limit cycle oscillation frequencies are equivalent to laser oscillation frequencies, synchronization can be directly confirmed by the number of emission peaks in a measured spectrum. Namely, if a spectrum has a single emission peak, two lasers are synchronized, while if there are two emission peaks, they are not. Although coupling strengths between cavities are usually fixed in a device, it is still possible to actively tune the resonance frequency of a cold cavity \cite{Chong2004,Fushman2007,Tanabe2009,Faraon2009,Yuece2018} and effectively change coupling strengths as shown in Fig. \ref{fig:sync_lasers}(d). In this context, the spectral shape of laser emissions will be of interest. Below the lasing threshold, since laser cavities will behave as ``cold cavities", their emission spectrum is expected to exhibit normal-mode splitting. Meanwhile, above the lasing threshold the emission spectrum exhibits a single peak due to synchronization. Therefore, we may prove synchronization from the pump-power dependence of the change in spectral shape . Another promising experimental strategy to prove synchronization may be to pump two lasers independently and tune the respective laser frequencies by making use of the carrier-induced blue shift \cite{Takiguchi2016}. This strategy can be easily realized with spatially separated optical pumps or two electrodes for electric pumping.

The buried MQW PhC laser technique in the design of a real device is reported in Refs. \cite{Takeda2013,Matsuo2013}, where the PhC slab and buried PhC are composed of InP and InGaAsP/InGaAs, respectively. Furthermore, buried MQW PhC lasers can be pumped optically or electrically. If the photon lifetime of buried MQW PhC lasers is assumed to be $1/\gamma_1=1$ ps ($\sim$160 GHz), the frequency difference between two lasers corresponding to $\Delta\omega=0.01\gamma_1$, which is assumed in the simulations in Fig. \ref{fig:sync_lasers}, is $\Delta\omega=0.01\gamma_1\sim1.6$ GHz. This laser frequency difference may seem to be severe for experimental realization (even with the state-of-the-art fabrication technology, the frequency difference between cavities may be about 50 GHz \cite{Taguchi2011}), but we found that, qualitatively, the same synchronization can occur for a larger frequency difference. For example, synchronization with a laser frequency difference $\Delta\omega=0.1\gamma_1$ is discussed in Section 3 in the SM. Furthermore, we found that even if all the parameters of the three cavities including $\beta_{1,2}$ are moderately different, synchronization can occur (not shown).

\section*{Conclusion and outlook}
To conclude, we theoretically proposed a design of indirectly-coupled PhC cavity lasers that emulates the local Kuramoto model. In this study, we reinterpreted the injection-locking phenomenon of lasers as the synchronization of limit cycle oscillations. Furthermore, our design prevents laser oscillations from forming normal-modes (strong-coupling) with indirect coupling via additional cold cavities and realizes effective dissipative coupling without time-delay. Experimentally, this proposed structure will best be realized best by using buried MQW PhC cavities. First, after modelling laser oscillation with the Stuart-landau equation, we numerically demonstrated the synchronization of two indirectly-coupled PhC lasers using the coupled-mode equations of motion. Second, by applying the phase reduction theory to the two indirectly coupled lasers, we obtained corresponding phase equations of motion, which are equivalent to the local Kuramoto model. Finally, we briefly discussed synchronization dynamics for a one-dimensional chain of indirectly coupled PhC lasers and demonstrated that the proposed device can actually emulate the local Kuramoto chain. 

For future perspectives, first of all, the one-dimensional local Kuramoto model briefly investigated in Section 5, already comprises rich physics that were actively investigated by detailed numerical simulations \cite{Zheng1998} and renormalization group analysis \cite{Daido1988,Kogan2009}. Furthermore, very recently, Ref. \cite{Waechtler2020} demonstrated that even topological phenomena emerge in the one-dimensional chain of limit cycles. Thanks to the scalability of PhC cavities, the extension of the one-dimensional chain of PhC lasers to a two-dimensional array is straightforward, which is the realization of the celebrated two-dimensional local Kuramoto model \cite{Sakaguchi1987,Daido1988,Strogatz1988,Hong2005,Acebron2005,Rogister2007,Lee2010}. Compared with the in-phase synchronization case, large-scale anti-phase synchronization has not yet been drawing attention. For instance, as Ref. \cite{VathakkattilJoseph2020} indicates that a large anti-phase synchronization network is not possible, large-scale anti-phase synchronization itself may be of fundamental interest. Another important direction will be the inclusions of classical and quantum noise effects in indirectly coupled PhC lasers, which will provide spectral information. As we briefly discussed in Section 6, we may prove synchronization in terms of the pump power dependence of spectral shape. In this direction, it is also easy to construct a quantum model corresponding to our coupled-mode equations. In fact, the quantum counterpart of the classical Stuart-Landau model is the Scully-lamb master equation \cite{Scully1967,Takemura2019a}. Therefore, even the effect of quantum noises on synchronization \cite{Vinokur2008,Lee2013,Walter2014} may be tested with the proposed device. Moreover, since the synchronization problem in the local Kuramoto model is analogous to the energy minimization problem in the XY model, our device may be used for simulating the spin system in statistical physics \cite{Flovik2016,VathakkattilJoseph2020}. Finally, from the standpoint of practical application, an injection-locked (synchronized) PhC laser array can be employed as a single-mode high-power PhC laser. Even though every PhC laser unavoidably has a different oscillation frequency, in the fully synchronized state, they behave as a laser with a single frequency. Furthermore, this type of a laser will also have high coherence because all laser phases are locked in the synchronized state. 

\section*{Methods}
All the time evolutions were obtained by integrating the coupled-mode equations of motion with the conventional Runge-Kutta method. The synchronization trees were calculated as the mean oscillation frequencies of time evolutions. The calculation of the mean frequencies is based on the peak detection technique. To precisely determine the mean frequencies, long time  evolutions (typically 6000$\gamma_1^{-1}$) were required. 


%

\section*{Author contributions statement}
N.T. developed the main theoretical idea and performed the simulations. K. T. taught us the injection locking of lasers and coupled-mode theory. K. T., M. T., and M. N. contributed to the discussion on photonic crystal laser technologies. N. T. drafted the manuscript. 


\section*{Competing interests}
The authors declare no competing interests.

\clearpage

\renewcommand{\thefigure}{S\arabic{figure}}
\setcounter{figure}{0}
\renewcommand{\theequation}{S\arabic{equation}}
\setcounter{equation}{0}
\renewcommand{\thetable}{S\arabic{table}}
\setcounter{table}{0}

\section{Derivation of phase equations of motion}
We detail the calculations for the phase reduction analysis discussed in Section 4.2. Note that here we do not describe the phase reduction theory itself, which is explained, for example, in a review \cite{Nakao2016} and textbook \cite{Kuramoto2003}. First of all, we consider an orbit for laser L1 given by
\begin{eqnarray}
\left( \begin{array}{c}
x(\phi)\\
y(\phi)
\end{array} \right)
=\sqrt{\frac{\epsilon_1}{\beta_1}}
\left( \begin{array}{c}
-\cos\phi\\
\sin\phi
\end{array} \right),
\label{eq:orbit_S}
\end{eqnarray} 
where $\phi=\omega_1t$. Additionally, $x(\phi)$ and $y(\phi)$ represent real and imaginary parts of the laser field, respectively. Importantly, here, we consider the laser oscillation L1  given by Eq. (\ref{eq:orbit_S}) as a ``standard oscillator" for the phase reduction analysis. In the phase reduction theory, the function ${\bm Z}(\phi)$ called ``sensitivity", which represents the response of limit cycle oscillation to perturbation, plays a central role. Fortunately, the analytical expression of the sensitivity ${\bm Z}(\phi)$ for the Stuart-Landau model [see Eq. (3) in the main text] is known \cite{Nakao2016} and, for the orbit given by Eq. (\ref{eq:orbit_S}), is
\begin{eqnarray}
{\bm Z}(\phi)=
\left( \begin{array}{c}
Z_x(\phi)\\
Z_y(\phi)
\end{array} \right)
=\sqrt{\frac{\beta_1}{\epsilon_1}}
\left( \begin{array}{c}
\sin\phi\\
\cos\phi
\end{array} \right).
\end{eqnarray}
Our goal is to obtain the phase equations of motion in the form 
\begin{eqnarray}
\dot{\phi_1}&=&\omega_0+\delta\Omega_1+\Gamma_{12}(\phi_1-\phi_2)\label{eq:phase_eq_s1}\\
\dot{\phi_2}&=&\omega_0+\delta\Omega_2+\Gamma_{21}(\phi_2-\phi_1),\label{eq:phase_eq_s2}
\end{eqnarray}
where $\phi_{1}$ and $\phi_{2}$ are the phases of laser L1 and L2, respectively. Meanwhile, $\omega_0$ is the oscillation frequency of the standard oscillator. Here, $\delta\omega_{1,2}$ represents a frequency shift originating from the difference between the standard oscillator and laser oscillation L1 and L2.  Meanwhile, $\Gamma_{12}(\psi)$ and $\Gamma_{21}(\psi)$ are the phase coupling functions. We calculate $\delta\omega_{1,2}(\theta)$ and $\Gamma_{ij}(\psi)$ using the following formulae: 
\begin{equation}
\delta\omega_{1,2}=\frac{1}{2\pi}\int_0^{2\pi}d\theta{\bm Z}(\theta)\cdot\delta{\bm F}_{1,2}(\theta)
\label{eq:deltaw}
\end{equation}
and
\begin{equation}
\Gamma_{ij}(\psi)=\frac{1}{2\pi}\int_0^{2\pi}d\eta{\bm Z}(\eta+\psi)\cdot{\bm G}_{ij}(\eta),\label{eq:Gamma_int}
\end{equation}
where $\delta{\bm F}_{1,2}(\theta)$ represents the difference between the standard oscillator and laser oscillation  L1 and L2. Since, here, the standard oscillator is nothing else but laser L1, $\delta{\bm F}_{1}(\theta)=0$ holds, and, consequently, we find that $\omega_0=-\omega_1$ and $\delta\omega_{1}=0$. For the approximated equations of motion (10) and (11) in the main text, for the orbit Eq. (\ref{eq:orbit_S}), the terms $\delta{\bm F}_{2}(\theta)$ and ${\bm G}_{12}(\eta)$ are represented as
\begin{eqnarray}
\delta{\bm F}_{2}(\theta)
=-\Delta\omega\sqrt{\frac{\epsilon_1}{\beta_1}}
\left( \begin{array}{c}
\sin\theta\\
\cos\theta
\end{array} \right)
\end{eqnarray}
and 
\begin{eqnarray}
{\bm G}_{12}(\eta)
=
-\frac{2g_1g_2}{\Gamma_1}
\sqrt{\frac{\epsilon_1}{\beta_1}}
\left( \begin{array}{c}
-\cos\eta\\
\sin\eta
\end{array} \right).
\end{eqnarray}
Now, Eqs (\ref{eq:deltaw}) and (\ref{eq:Gamma_int}) are easily calculated as 
\begin{equation}
\delta\omega_{2}=-\frac{1}{2\pi}\int_0^{2\pi}d\theta\Delta\omega(\sin^2\theta+\cos^2\theta)=-\Delta\omega=\omega_1-\omega_2
\end{equation}
and
\begin{eqnarray}
\Gamma_{12}(\psi)=\Gamma_{21}(\psi)&=&\frac{2g_1g_2}{\Gamma_1}\frac{1}{2\pi}\int_0^{2\pi}d\eta\left\lbrace\sin(\eta+\psi)\cos\eta-\cos(\eta+\psi)\sin\eta\right\rbrace\nonumber\\
&=&\frac{2g_1g_2}{\Gamma_1}\frac{1}{2\pi}\int_0^{2\pi}d\eta \sin\psi=\frac{2g_1g_2}{\Gamma_1}\sin\psi.
\end{eqnarray}
Therefore, as discussed in the main text, if the adiabatic elimination approximation is valid, for the approximated coupled-mode equations, we obtain the corresponding phase equations of motion (17) and (18) in the main text.

\section{Synchronization dynamics with a large coupling strength}
\begin{figure}
\includegraphics[width=0.9\textwidth]{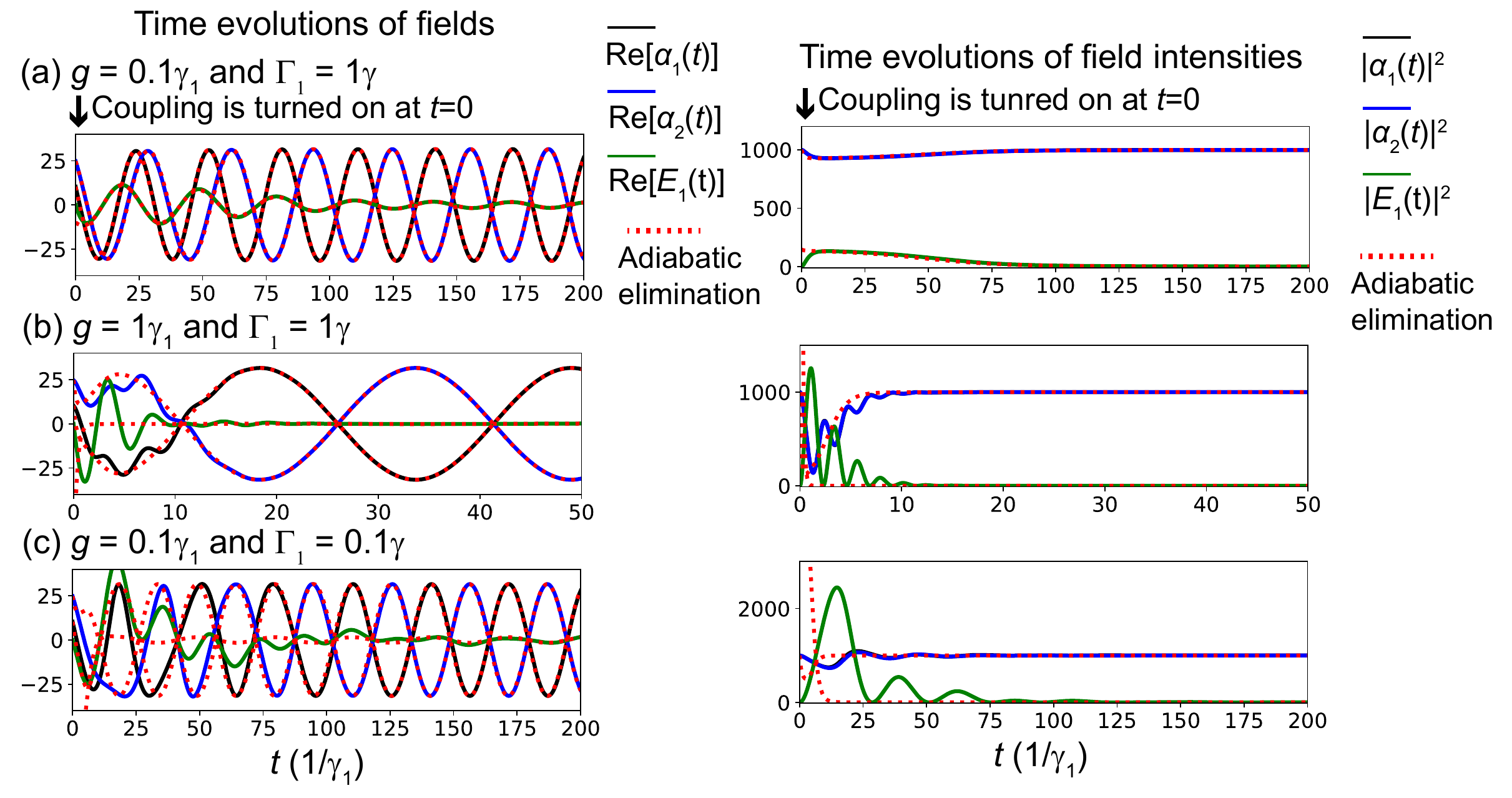}
\caption{Synchronization dynamics are simulated by turning on coupling at $t=0$. The time evolutions of the fields ${\rm Re}[\alpha_{1,2}(t)]$ and ${\rm Re}[E(t)_{1}]$ are shown in the left panel, while field intensities $|\alpha_{1,2}(t)|^2$ and ${\rm Re}|E(t)_{1}|^2$ are shown in the right panel. The solid curves are simulations calculated with the original coupled-mode equations (6)-(8) in the main text, while the dashed curves are simulations obtained with the adiabatic elimination approximation [Eqs. (10) and (11) in the main text]. On the left panels, the shifted frequencies of the lasers are $\omega_1^\prime=0.2\gamma_1$ and $\omega_2^\prime=0.21\gamma_1$. The parameters used in (a) are the same as those in Fig. 3(a) in the main text. On the other hand, we used $g_1=g_2=1\gamma_1$ and $\Gamma_1=1\gamma_1$ for (b), while we used $g_1=g_2=0.1\gamma_1$ and $\Gamma_1=0.1\gamma_1$ for (c). The other parameters used in (b) and (c) are the same as those in Fig. 3(a) in the main text.}
\label{fig:validity}
\end{figure}

Here, by simulating synchronization dynamics for two indirectly coupled lasers, we discuss the validity of the adiabatic elimination approximation when coupling strengths are larger than field decay rates. In the left panels of Fig. \ref{fig:validity}, in the same way as in Fig. 3(a) in the main text, we show the time evolutions of the fields ${\rm Re}[\alpha_{1,2}(t)]$ and ${\rm Re}[E_{1}(t)]$ calculated with the original equations [see Eqs. (6)-(8) in the main text] (solid lines) and calculated with the approximated equations [see Eqs. (10) and (11) in the main text] (dashed lines). Meanwhile, in the right panels of Fig. \ref{fig:validity}, we show the time evolutions of field intensities $|\alpha_{1,2}(t)|^2$ and $|E_{1}(t)|^2$ calculated with the original equations (solid lines) and calculated with the approximated equations (dashed lines). Importantly, in the same way as in Fig. 3(a) in the main text, since coupling is switched on for uncoupled laser oscillations at $t=0$, Fig. \ref{fig:validity} represents synchronization dynamics. Furthermore, to clearly show synchronization dynamics, we set the shifted frequencies as $\omega_1^\prime=0.2\gamma_1$, $\omega_2^\prime=0.21\gamma_1$, and $\Omega_1^\prime=0.2\gamma_1$, which are the same as in Fig. 3(a).

For the all simulations in Fig. \ref{fig:validity}, we used the same parameters as in Fig. 3(a) in the main text except for the coupling strength $g_{1,2}$ and the decay rate of the cold cavity $\Gamma_1$. First, Fig. \ref{fig:validity}(a) shows the simulations with $g_1=g_2=0.1\gamma_1$ and $\Gamma_1=1\gamma_1$. Therefore, the left panel in Fig. \ref{fig:validity}(a) is the same as Fig. 3(a) in the main text. Both the left and right panels in Fig. \ref{fig:validity}(a) indicate that the adiabatic elimination approximation perfectly reproduces synchronization dynamics calculated with the original coupled-mode equations [compare the solid and dashed curves in Fig. \ref{fig:validity}(a)]. Note that there is no oscillation in the field intensity dynamics in the right panel in \ref{fig:validity}(a). Second, for the simulation in Fig. \ref{fig:validity}(b), we increased the coupling strength as $g_{1,2}=1\gamma_1$, while we fixed the decay rate of the cold-cavity as $\Gamma_1=1\gamma_1$. As the left panel in Fig. \ref{fig:validity}(b) indicates, the adiabatic elimination approximation fails to reproduce the turn-on dynamics (compare the solid and dashed curves until $t\simeq10\gamma_1^{-1}$). Meanwhile, the adiabatic elimination approximation succeeds in reproducing the ``steady-state" synchronized laser oscillations (compare the solid and dashed curves after $t\simeq10\gamma_1^{-1}$). The right panel in Fig. \ref{fig:validity}(b) indicates that the failure of the adiabatic elimination approximation is associated with the coherent oscillations of field intensities between the cavities in the turn-on dynamics, which originate  from the (near) strong-coupling condition. Furthermore, as Fig. \ref{fig:validity}(b) shows, it is when the coherent intensity oscillations are damped and the system reaches the ``steady-state" around $t\simeq10\gamma_1^{-1}$ that the adiabatic elimination approximation becomes valid. Finally, in Fig. \ref{fig:validity}(c), we show simulations with the smaller decay rate of the cold cavity $\Gamma_1=0.1\gamma_1$, but with a fixed coupling strength $g_{1,2}=0.1\gamma_1$. As in Fig. \ref{fig:validity}(b), the left panel in Fig. \ref{fig:validity}(c) indicates that the adiabatic elimination approximation cannot reproduce the turn-on dynamics (before $t\simeq100\gamma_1$), while the approximation starts to well approximate the ``steady-state" synchronized laser oscillations after $t\simeq100\gamma_1$. This result can again be explained in terms of the coherent oscillation of the field intensities originating from the (near) strong coupling condition [see the right panel in Fig. \ref{fig:validity}(c)].

In summary, we found that when coherent intensity oscillation is present in the strong coupling regime, the adiabatic elimination approximation fails to reproduce turn-on dynamics. Intuitively, since the adiabatic elimination approximation assumes a large time-scale difference between the field dynamics in the cold cavity (fast dynamics) and in the laser cavities (slow dynamics), when these fields exhibit coherent oscillations, the time-scale separation becomes impossible and, consequently, the adiabatic elimination approximation breaks down. Therefore, the strict conditions required for the adiabatic elimination approximation are $\omega_1\simeq\omega_2\simeq\Omega_1$ and $\ g_{1,2}<\gamma_{1,2},\Gamma_1$ [see Eq. (12) in the main text]. We also stress that even when the above conditions are not satisfied, the stable synchronization of laser oscillations itself can be realized, and the adiabatic elimination approximation well describes ``steady-state" synchronized laser oscillations as we find in Fig.\ref{fig:validity}(b) and (c), which is because the coherent intensity oscillation is damped in the ``steady-state" and the time-scale separation becomes possible. 
 
\section{Synchronization of lasers with a larger oscillation frequency difference}
\begin{figure}
\includegraphics[width=0.9\textwidth]{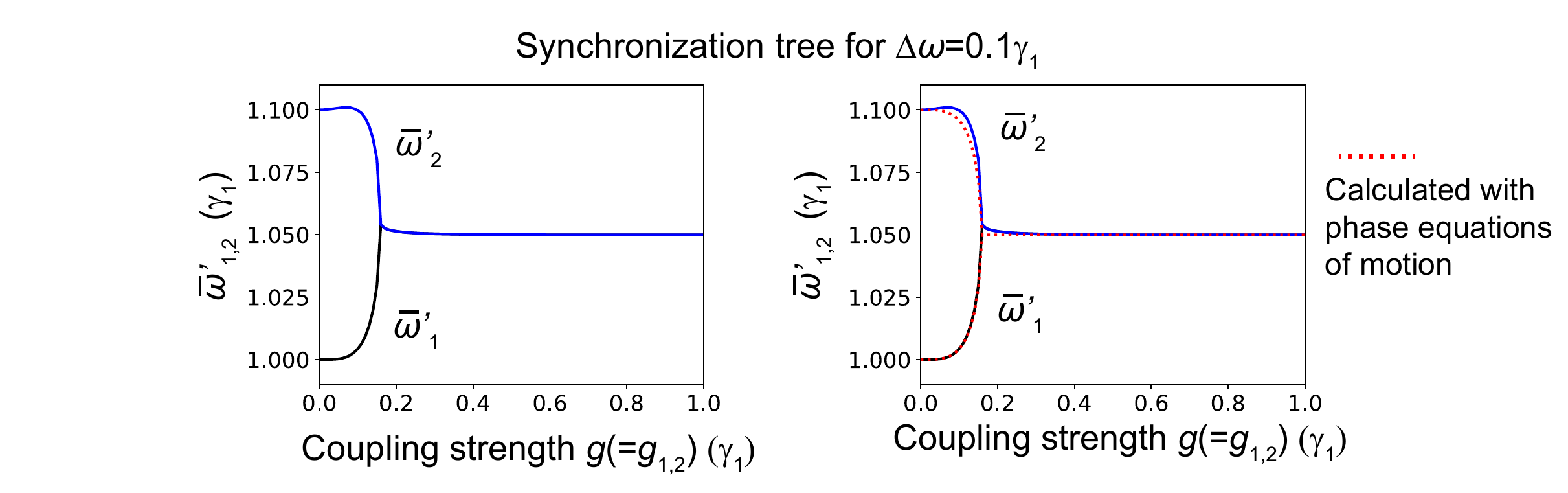}
\caption{Mean frequency of the oscillation frequency $\bar{\omega}_{1,2}$ of the laser L1 and L2 as a function of the coupled strength $g_{1,2}$. The solid black and blue curves in the left and right panels represent the synchronization trees calculated with the coupled-mode equations (6)-(8) in the main text. Meanwhile, the red dashed curves are the synchronization trees calculated with the corresponding phase equations of motion (16) and (17) in the main text.}
\label{fig:Appendix}
\end{figure}

Here, we discuss the synchronization of two indirectly coupled lasers with a frequency difference $\Delta\omega\equiv\omega_2-\omega_1=0.1\gamma_1$, which is ten times larger than that used in Section 3 in the main text. In fact, all the parameters except for $\omega_2$ and $g_{1,2}$ are the same as those used in Section 3. Therefore, the shifted resonance frequencies of the laser cavities and cold cavities are set as $\omega_1^\prime=1\gamma_1$, $\omega_2^\prime=1.1\gamma_1$, and $\Omega_1^\prime=1\gamma_1$. First, by simulating the coupled-mode equations (6)-(8) in the main text, we obtained the synchronization tree in the left panel in Fig. \ref{fig:Appendix}, where the average frequency of laser oscillation $\bar{\omega}_{1,2}^\prime$ is shown as a function of $g_{1,2}$. The left panel in Fig. \ref{fig:Appendix} indicates that synchronization can occur even with this laser frequency difference $\Delta\omega=0.1\gamma_1$. However, comparing Fig. \ref{fig:Appendix} with Fig. 2(c) in the main text, we can easily find that the synchronization is slightly asymmetric between $\bar{\omega}_1^\prime$ and $\bar{\omega}_2^\prime$, which is because there is a relatively large asymmetry in the resonance frequencies of the cavities as $\omega_1^\prime=1\gamma_1$, $\omega_2^\prime=1.1\gamma_1$, and $\Omega_1^\prime=1\gamma_1$.

Second, in the right panel in Fig. \ref{fig:Appendix}, the solid black and blue curves are the same as those in the left panel, but the red dashed curves are calculated with the corresponding phase equations of motion (16) and (17) in the main text. The corresponding phase equations of motion can qualitatively reproduce the original synchronization tree (compare the solid and dashed curves), the two synchronization trees do not perfectly coincide with each other. In fact, the ideal synchronization tree obtained with the phase equations of motion (see the dashed curves) cannot reproduce the asymmetry between $\bar{\omega}_1^\prime$ and $\bar{\omega}_2^\prime$ in the solid curves. This difference between the two approaches originates from the fact that the adiabatic elimination approximation partly fails because the adiabatic elimination condition $\omega_1\simeq\omega_2\simeq\Omega_1$ [see Eq. (12) in the main text] is partly violated, which results in decreasing the approximation of the phase equations of motion. 

Finally, we comment on the critical coupling strengths of synchronization, which is indicated both by the solid and dashed curves as $g_1=g_2=\sqrt{0.025}\gamma_1$. We show that the critical coupling strengths agree with the prediction of the phase equation of motion (15) in the main text. Namely, for $\Gamma_1=1\gamma_1$ and $\Delta\omega=0.1\gamma_1$, when $g_1=g_2=\sqrt{0.5}\gamma_1$, the synchronization condition $-4g_1g_2/\Gamma_1\leq\Delta\omega\leq4g_1g_2/\Gamma_1$ is satisfied. We also found that with a further increase in the coupling strength $g_{1,2}$, the adiabatic elimination approximation becomes better around $g_{1,2}=0.2\gamma_1$ and that both solid and dashed synchronization trees reach $\bar{\omega}_{1,2}^\prime\rightarrow1.05\gamma_1$, which is the mean frequency for $\omega_1^\prime=1\gamma_1$ and $\omega_2^\prime=1.1\gamma_1$ without coupling.

\section{Synchronization of class-B lasers}
\begin{figure}
\includegraphics[width=0.9\textwidth]{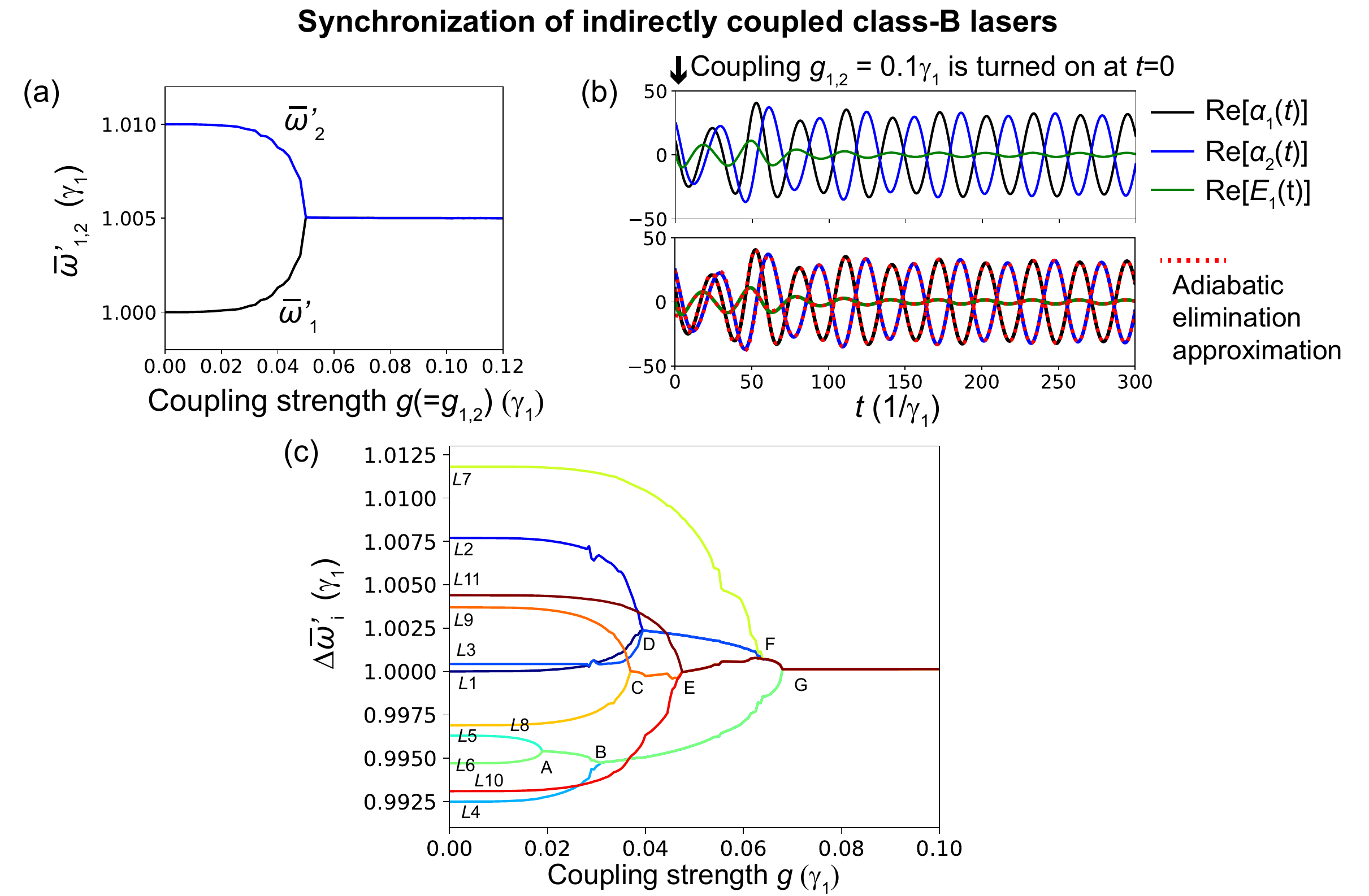}
\caption{(a) Mean frequencies of two indirectly coupled class-B lasers $\bar{\omega}_{1}$ and $\bar{\omega}_{2}$ as a function of coupled strength $g_{1,2}$. (b) Synchronization dynamics of two indirectly coupled class-B lasers, where coupling ($g_{1,2}=0.1\gamma_1$) is turned on at $t=0$. Time evolution of the laser fields ${\rm Re}[\alpha_{1,2}(t)]$ and ${\rm Re}[E(t)_{1}]$ are shown. The upper panel shows the time evolutions calculated with the original coupled-mode equations (\ref{eq:a1})-(\ref{eq:N2}). In the lower panel, the time evolutions calculated with the adiabatic elimination approximation Eqs. (\ref{eq:ad1})-(\ref{eq:ad2}) are plotted as red dashed curves on the original curves. The shifted frequencies of the laser and cold cavities $\omega_1^\prime=0.2\gamma_1$, $\omega_2^\prime=0.21\gamma_1$, and $\Omega_1^\prime=0.2\gamma_1$ are used. For the other parameters in (a) and (b), we used the same values as in Fig. 2(c) and 3(a) in the main text except for the carrier lifetimes $\tilde{\gamma}_2=\tilde{\gamma}_1=0.01\gamma_1$. (c) The mean oscillation frequencies of the eleven lasers $\bar{\omega}_i^\prime$ as a function the coupling strength $g_i=g$ for all $i$. The synchronization points are denoted by A-G. The parameter values other than  $\tilde{\gamma}_i=0.01\gamma_1$ are the same as those used in Fig. 4(b) in the main text.}
\label{fig:classB}
\end{figure}

In this section, we demonstrate that all the arguments in the manuscript based on the Stuart-Landau equation can be reproduced even quantitatively with class-B lasers where $\gamma_c>\gamma_\|$ [see Eqs. (1) and (2) in the main text]. Since the adiabatic elimination of the carrier degree of freedom cannot be allowed, we need to directly simulate the following coupled-mode equations for two indirectly coupled class-B lasers:
\begin{eqnarray}
\dot{\alpha}_1&=&-i\omega_1\alpha_1-\frac{1}{2}\gamma_1\alpha_1+\frac{1}{2}\beta_1\tilde{\gamma}_{1}N_1\alpha_1-ig_1E_1\label{eq:a1}\\
\dot{N}_1&=&-\tilde{\gamma}_1N_1-\beta\tilde{\gamma}_1N_1|\alpha_1|^2+P_1\label{eq:N1}\\
\dot{E}_1&=&-i\Omega_1E_1-\frac{1}{2}\Gamma_1E_1-ig_1\alpha_1-ig_2\alpha_2\label{eq:az}\\
\dot{\alpha}_2&=&-i\omega_2\alpha_2-\frac{1}{2}\gamma_2\alpha_2+\frac{1}{2}\beta_2\tilde{\gamma}_{2}N_2\alpha_2-ig_2E_1\label{eq:a1}\\
\dot{N}_2&=&-\tilde{\gamma}_2N_2-\beta_2\tilde{\gamma}_2N_2|\alpha_2|^2+P_2,\label{eq:N2}
\end{eqnarray}
where $N_{1,2}$ represents the carrier number for laser L1,2, while $\tilde{\gamma}_{1,2}$ is the decay rate of the carrier $N_{1,2}$. With Eq. (4) in the main text, the pump power $P_{i}$ is connected with the pump parameter $\epsilon_{i}$ as
\begin{equation}
P_i=\frac{\gamma_i}{\beta_i}(1+\epsilon_i).
\end{equation}
The definitions of the other parameters are the same as in Eqs (6)-(8) in the main text. For all simulations in this supplemental material, the photon and carrier lifetimes are set as $\tilde{\gamma}_{1}=\tilde{\gamma}_{2}=0.01\gamma_{1}$ and $\gamma_2=\gamma_1$, which is clearly the class-B regime. For the other parameters, we use the same values as in the main text. Namely, we use $\beta_1=\beta_2=0.001$ and $\varepsilon_1=\varepsilon_2=1.0$. For the frequencies of the laser and cold cavity, we use the shifted frequencies $\omega_1^\prime=1\gamma_1$, $\omega_2^\prime=1.01\gamma_1$, and $\Omega_1^\prime=1\gamma_1$. First, in Fig. \ref{fig:classB}(a), we show the mean frequencies of the two lasers $\bar{\omega}_{1}^\prime$ and $\bar{\omega}_{2}^\prime$ as a function of the coupling strength between cavities $g_{1,2}$. We found that Fig. \ref{fig:classB}(a) is even quantitatively the same as Fig. 2(c) in the main text.

The only difference between the class-A and class-B lasers is the response to amplitude perturbation. In the same way as in Fig. 3(a) in the main text, synchronization dynamics are shown in Fig. \ref{fig:classB}(b), where coupling ($g_{1,2}=1\gamma_1$) is turned on at $t=0$. The turn-on of coupling works as perturbation and induces relaxation oscillations in the synchronization dynamics [see the oscillations of the amplitudes in Fig. \ref{fig:classB}(b)], which is the characteristic of class-B lasers. However, after the relaxation oscillations are damped, the two class-B laser oscillations exhibit anti-phase synchronization, which is the same as class-A lasers. Additionally, in the lower panel in Fig. \ref{fig:classB}(b), the red dashed curves represent approximated equations of motion
\begin{eqnarray}
\dot{\alpha}_1&=&-i\omega_1\alpha_1-\frac{1}{2}\gamma_1\alpha_1-\frac{2g_1^2}{\Gamma_1}\alpha_1+\frac{1}{2}\beta_1\tilde{\gamma}_{1}N_1\alpha_1-\frac{2g_1g_2}{\Gamma_1}\alpha_2\label{eq:ad1}\\
\dot{N}_1&=&-\tilde{\gamma}_1N_1-\beta\tilde{\gamma}_1N_1|\alpha_1|^2+P_1\\
\dot{\alpha}_2&=&-i\omega_2\alpha_2-\frac{1}{2}\gamma_2\alpha_2-\frac{2g_2^2}{\Gamma_1}\alpha_2+\frac{1}{2}\beta_2\tilde{\gamma}_{2}N_2\alpha_2-\frac{2g_1g_2}{\Gamma_1}\alpha_1\\
\dot{N}_2&=&-\tilde{\gamma}_2N_2-\beta_2\tilde{\gamma}_2N_2|\alpha_2|^2+P_2,\label{eq:ad2}
\end{eqnarray}
which are obtained by adiabatically eliminating the field of the cold-cavity $E_1$ with Eq. (9) in the main text. The lower panel in Fig. \ref{fig:classB}(b) clearly indicates that the adiabatic elimination of the field $E_1$ is a very good approximation even for class-B lasers. By applying the numerical phase reduction to the rate equations of motion (1) and (2) in the main text\cite{Nakao2016,Takemura2020a}, we found that the sensitivity ${\bm Z}(\phi)$ (see Section 1 in this supplemental material) is given by ${\bm Z}(\phi)=(Z_x(\phi),Z_y(\phi)),Z_N(\phi))=\sqrt{\beta_1/\epsilon_1}(-\cos\phi,\sin\phi,0)$. Thus, the phase equations of motion corresponding to Eqs. (\ref{eq:ad1})-(\ref{eq:ad2}) are the same as Eqs (17) and (18) in the main text. Accordingly, the antisymmetric part of the phase coupling function is also given by $\Gamma_a(\psi)=(4g_1g_2/\Gamma_1)\sin\psi$, and phase locking occurs at the phase $\psi=\phi_2-\phi_1=\pi$ (anti-phase synchronization). Since the phase equations of motion are the same between class-A and class-B lasers, all the arguments on phase dynamics for class-A lasers can be applied to class-B lasers. 

Finally, in Fig. \ref{fig:classB}(c), we show a synchronization tree  for eleven indirectly coupled class-B lasers. All the parameters except for carrier lifetime $\tilde{\gamma}_i=0.01\gamma_1$, which is not present in the main text, are the same as those in Fig. 4(b) in the main text. Figure \ref{fig:classB}(c) is again quantitatively the same as the synchronization tree shown in Fig. 4(b) in the main text.

\section{Impact of the linewidth enhancement factor on synchronization}
In this section, we briefly discuss the impacts of the linewidth enhancement factor (the Henry factor \cite{Henry}) $\alpha_{\rm H}$ on synchronization of indirectly coupled lasers. Since the linewidth enhancement factor is not negligible  in semiconductor lasers, its effects will be very important for real experiments using PhC lasers. In the same way as Refs. \cite{Henry,Winful1988,Hamel2015}, we introduce the linewidth enhancement factor $\alpha_{\rm H}$ into rate equations (1) and (2) in the main text as
\begin{eqnarray}
\dot{\alpha}&=&-i\omega_c\alpha-\frac{1}{2}\gamma_c\alpha+(1-i\alpha_{\rm H})\frac{1}{2}\beta\gamma_\|N\alpha\label{eq:rate_alpha}\\
\dot{N}&=&-\gamma_\|N-\beta\gamma_\|N|\alpha|^2+P.\label{eq:rate_N}
\end{eqnarray}
The above modified rate equations indicate that the factor $\alpha_{\rm H}$ contributes to a carrier-induced blue shift. In this section, we set the value of the linewidth enhancement factor $\alpha_{\rm H}=4.0$, which is the measured value for buried multiple quantum well PhC lasers \cite{Kim2012}. For indirectly coupled lasers coupled-mode equations including the linewidth enhancement effect  are explicitly written as 
\begin{eqnarray}
\dot{\alpha}_1&=&-i\omega_1\alpha_1-\frac{1}{2}\gamma_1\alpha_1+(1-i\alpha_{\rm H})\frac{1}{2}\beta_1\tilde{\gamma}_{1}N_1\alpha_1-ig_1E_1\label{eq:a1H}\\
\dot{N}_1&=&-\tilde{\gamma}_1N_1-\beta\tilde{\gamma}_1N_1|\alpha_1|^2+P_1\\
\dot{E}_1&=&-i\Omega_1E_1-\frac{1}{2}\Gamma_1E_1-ig_1\alpha_1-ig_2\alpha_2\\
\dot{\alpha}_2&=&-i\omega_2\alpha_2-\frac{1}{2}\gamma_2\alpha_2+(1-i\alpha_{\rm H})\frac{1}{2}\beta_2\tilde{\gamma}_{2}N_2\alpha_2-ig_2E_1\\
\dot{N}_2&=&-\tilde{\gamma}_2N_2-\beta_2\tilde{\gamma}_2N_2|\alpha_2|^2+P_2,\label{eq:N2H}
\end{eqnarray}
where all the parameters other than $\alpha_{\rm H}$ are already defined in Section 4 in this supplemental material.

\begin{figure}
\includegraphics[width=0.9\textwidth]{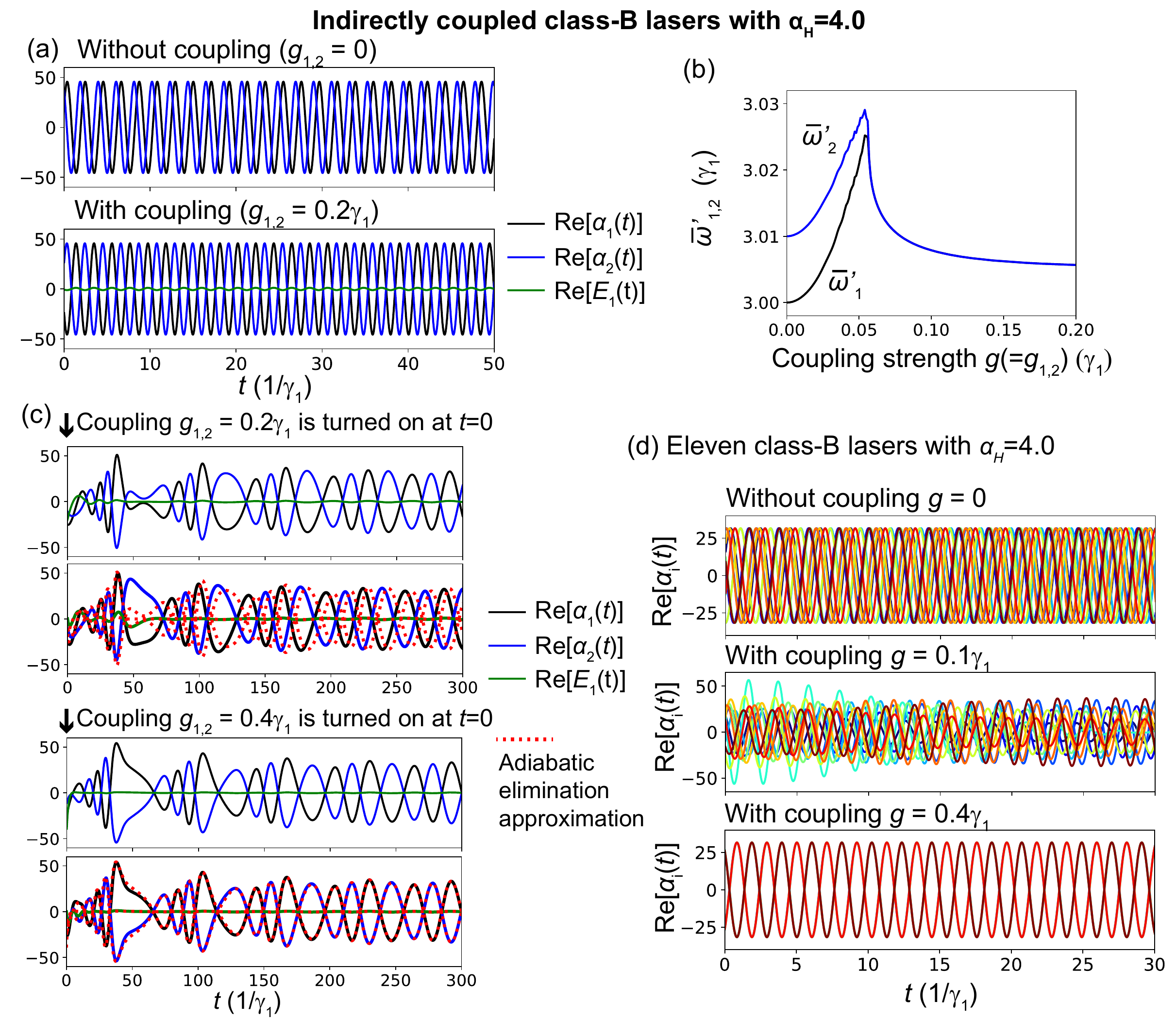}
\caption{Simulations with the linewidth enhancement parameter $\alpha_{\rm H}=4.0$. (a) Time evolutions of the real part of the field ${\rm Re}[\alpha_{1,2}(t)]$ without $g_{1,2}=0$ and with coupling $g_{1,2}=0.2\gamma_1$. The shifted cavity frequencies are $\omega_1^\prime=1\gamma_1$ and $\omega_2^\prime=1.01\gamma_1$, and $\Omega_1^\prime=3\gamma_1$. (b) Mean frequency of the laser oscillation $\bar{\omega}_{1,2}$ as a function of coupled strength $g_{1,2}$. (c) Synchronization dynamics of the fields ${\rm Re}[\alpha_{1,2}(t)]$ and ${\rm Re}[E(t)_{1}]$. Coupling ($g_{1,2}=0.2\gamma_1$ and $0.4\gamma_1$) is turned on at $t=0$. The solid and dashed curves represent synchronization dynamics calculated with the original and approximated equations of motion, respectively. Here, the cavity frequencies are shifted as $\omega_1^\prime=-1.8\gamma_1$ and $\omega_2^\prime=-1.79\gamma_1$, and $\Omega_1^\prime=0.2\gamma_1$. For the other parameters in (a), (b), and (c), we used $\beta_2=\beta_1=0.001$, $\varepsilon_2=\varepsilon_1=1.0$, $\gamma_2=\gamma_1\equiv1$, $\tilde{\gamma}_2=\tilde{\gamma}_1=0.01\gamma_1$, and $\Gamma_1=1\gamma_1$. (d) Time evolutions of the real parts of the eleven indirectly coupled laser fields for $g(=g_i)=0$ (top), $0.1\gamma_1$ (middle), and $0.4\gamma_1$ (bottom). The parameter values except for $\tilde{\gamma}_i=0.01\gamma_1$, $\alpha_{\rm H}=4.0$, and $\Omega_i=3\gamma_1$ are the same as those used in Fig. 4(b) in the main text.}
\label{fig:Henry}
\end{figure}

Figure \ref{fig:Henry}(a) shows the time evolutions of two indirectly coupled lasers calculated with Eqs. (\ref{eq:a1H})-(\ref{eq:N2H}). For the parameters, we use the carrier lifetime $\tilde{\gamma}_{1}=\tilde{\gamma}_{2}=0.01\gamma_1$, linewidth enhancement factor $\alpha_{\rm H}=4.0$, and $\Omega_1^\prime=3\gamma_1$. The other parameters are the same values as in Fig. 2(a) in the main text: $\omega_1^\prime=1\gamma_1$, $\omega_2^\prime=1.01\gamma_1$, $\beta_1=\beta_2=0.001$, $\varepsilon_2=\varepsilon_1=1.0$, and $\gamma_2=\gamma_1\equiv1$. Later, we explain the reason why the frequency of the cold cavity is very different from those of laser cavities ($\Omega_1^\prime=3\gamma_1$), which actually plays a key role in the synchronization of lasers with the linewidth enhancement factor . First, we discuss how the factor $\alpha_{\rm H}$ modifies independent laser oscillations . The upper panel in Fig. \ref{fig:Henry}(a) represents the time evolutions of the real parts of the fields ${\rm Re}[\alpha_{1}(t)]$ and ${\rm Re}[\alpha_{2}(t)]$ without coupling $g_{1,2}=0$. We notice that the laser oscillation frequencies are much higher than those in Fig. 2(a) in the main text. In fact, the laser oscillation frequencies are found to be $\bar{\omega}_1^\prime=3\gamma_1$ and $\bar{\omega}_2^\prime=3.01\gamma_1$ for lasers L1 and L2, respectively. The increases in the laser oscillation frequencies originate from the carrier-induced blue shift associated with $\alpha_{\rm H}$. With $\alpha_{\rm H}$, the oscillation frequency of a laser is shifted as $\omega_c+\alpha_H\beta\gamma_\|N/2$. Since the saturated carrier number above the lasing threshold is $N=\gamma_c/(\beta\gamma_\|)$, the frequency shift of laser oscillation is $\alpha\gamma_c/2$. In our case, the laser frequency shift is estimated as $\alpha\gamma_c/2=2\gamma_1$, which coincides with the simulation. Second, we introduce coupling $g_{1,2}=0.2\gamma_1$ in the lower panel of Fig. \ref{fig:Henry}(a), which shows that anti-phase synchronization can occur even with the presence of the linewidth enhancement factor $\alpha_{\rm H}=4.0$.

Now, we show the mean frequencies of the two lasers $\bar{\omega}_{1}^\prime$ and $\bar{\omega}_{2}^\prime$ as a function of the coupling strength $g_{1,2}$ in Fig. \ref{fig:classB}(b). In stark contrast to Fig. 2(c) in the main text, the synchronization tree shown in Fig. \ref{fig:classB}(b) is asymmetric, which originates from the modulations of laser frequencies induced by the changes in the carrier numbers. Although the changes in the carrier numbers are present even in Fig. 2(c) in the main text, they do not contribute to frequency shifts because $\alpha_{\rm H}=0$ in the main text. Figure 2(b) indicates that synchronization occurs at $g_{1,2}=0.056\gamma_1$ and, with a further increase in the coupling strength, the synchronized laser frequencies gradually decrease to $\bar{\omega}_{1}^\prime=\bar{\omega}_{2}^\prime=3.005\gamma_1$, which is the mean frequency of the two uncoupled lasers. 

Now, we explain the importance of the cold-cavity frequency for synchronization of lasers with the linewidth enhancement factor.  In Fig. \ref{fig:Henry}, we set the frequency of the cold cavity as $\Omega_1=3\gamma_1$ to compensate the carrier-induced blue shift induced by the factor $\alpha_{\rm H}$. With this trick, the adiabatic elimination condition for cavity frequencies [see Eq. (12) in the main text] effectively holds as $\omega_1+2\gamma_1\simeq\omega_2+2\gamma_1\simeq\Omega_1$, and the  adiabatic elimination of the cold cavity field leads to effective dissipative coupling. To check the validity of the adiabatic elimination approximation, we use Fig. \ref{fig:Henry}(c) and (d). Like Fig. 3(a) in the main text, Fig. \ref{fig:Henry}(c) and (d) show synchronization dynamics calculated with the original equations of motion (\ref{eq:a1H})-(\ref{eq:N2H}) (black and blue solid curves) and approximated equations of motion obtained with the adiabatic elimination of the field $E_1$ (red dashed curves). When the coupling strength is $g_{1,2}=0.2\gamma_1$, the approximated equations of motion can reproduce the frequencies of lasers but cannot reproduce their phases. Meanwhile, when the coupling strength is further increased to $g_{1,2}=0.4\gamma_1$, the approximated equations of motion can reproduce both frequencies and phases of the lasers. We also note that if the frequency of the cold cavity remains as $\Omega_1=1\gamma_1$, the adiabatic elimination fails and the laser oscillation becomes chaotic for a certain range of $g_{1,2}$ (not shown). This chaos emission of injection-locked lasers induced by $\alpha_{\rm H}$ is studied in Refs. \cite{Winful1988,Wang1988,Hwang2000}. Although the laser chaos is beyond the scope of this paper, this regime will be of great interest because our proposed device can also be used as an on-chip chaotic light emitter. 

Finally, we demonstrate that large-scale synchronization can be possible even with a non-negligible linewidth enhancement factor. The time evolutions of eleven indirectly coupled lasers with linewidth enhancement factors are shown in Fig. \ref{fig:Henry}(d), where we used the carrier lifetime $\tilde{\gamma}_i=0.01\gamma_1$ and $\Omega_i=3\gamma_1$, while all the laser cavities have $\alpha_{\rm H}=4.0$. The other parameters are the same as those in Fig. 4(b) in the main text. The upper, middle, lower panels in Fig. \ref{fig:Henry}(d) represent the time evolutions for $g_i=0$, $0.1\gamma_1$, and $0.4\gamma_1$, respectively. Note that, with $\alpha_{\rm H}=4.0$, unfortunately, we cannot show a synchronization tree like the one in Fig. 4(b) in the main text. This is because chaotic laser oscillations emerge for a certain coupling strength [for example, see $g=0.1\gamma_1$ in Fig. 4(d)]. On the other hand, when the coupling strength reaches a threshold, fully anti-phase synchronized oscillations can be realized even with $\alpha_{\rm H}=4.0$ [see $g=0.4\gamma_1$ in Fig. \ref{fig:Henry}(d)], which is an important indication for real experiments. Qualitatively, we may summarize the behavior of indirectly coupled laser with $\alpha_{\rm H}=4.0$ as follows. First, when non-zero coupling is introduced, laser oscillations are quasiperiodic or chaotic. It is not clear whether or not there is a threshold coupling strength of chaos transition as in Ref. \cite{Winful1988}. Second, as coupling strength is increased, the chaotic behavior is enhanced as shown in Fig. 4(d) for $g=0.1\gamma_1$. Finally, as the coupling strength is further increased, the chaotic laser oscillations suddenly exhibit anti-phase synchronization, which is the synchronization transition. For the parameters used in Fig. 4(d), the synchronization transition occur around $g(=g_i)=0.395\gamma_1$.

\section{Large-scale synchronization with disordered parameters}
Here, we discuss the large-scale synchronization of eleven indirectly coupled lasers but with disordered parameters. In Fig. 4(b) in the main text, for simplicity, we assumed that all cavities have the same parameter values except for laser cavity frequencies. However, this assumption is unrealistic because the parameters of all laser and cold cavities unavoidably have different values. Therefore, it is important to show the possibility of the large-scale synchronization of indirectly coupled lasers with disordered parameters. Note that, in this section, our objective is not to quantitatively investigate the upper limit of the disorder of parameters  for synchronization, but is only to demonstrate that large-scale synchronization can be possible even when parameter values are not the same. 
\begin{table}[htbp]
\caption{Parameter values for eleven laser and ten cold cavities}  
 \centering
 \begin{tabular}{l c c c c c r}
 \\ \hline\hline
 Index $i$ & $\omega_i^\prime$ [$\gamma_1$]& $\gamma_i$ [$\gamma_1$] & $\beta_i$ & $\epsilon_i$ & $\Omega_i$\ \ [$\gamma_1$] & $\Gamma_i$ [$\gamma_1$]\\ 
 \hline
 1 & 1.0000 & 1 & 0.0011 & 1.022 & 0.9906 & 0.9968\\
 2 & 1.0077 & 1.12 & 0.0012 & 0.923 & 0.9988 & 1.0068\\
 3 & 1.0004 & 1.11 & 0.00098 & 0.89 & 1.0011 & 1.0011\\
 4 & 0.9925 & 1.025 & 0.00099 & 1.035 & 0.9965 & 1.0035\\
 5 & 0.9963 & 0.962 & 0.0011 & 0.99 & 1.0220 & 0.9882\\
 6 & 0.9947 & 0.9977 & 0.00096 & 1.053 & 1.0032 & 1.0053\\
 7 & 1.0118 & 0.912 & 0.0012 & 0.979 & 0.9983 & 0.9979\\
 8 & 0.9969 & 1.09 & 0.0013 & 1.061 & 0.9974 & 1.0061\\
 9 & 1.0037 & 0.999 & 0.0012 & 0.983 & 1.0072 & 0.9983\\
 10 & 0.9931 & 1.0076 & 0.00095 & 0.951 & 0.9901 & 1.0049\\
 11 & 1.0044 & 1.033 & 0.00096 & 1.036 & & \\

 \hline\hline
 \end{tabular}
 \label{table1}
\end{table}

\begin{figure}[H]
\includegraphics[width=0.9\textwidth]{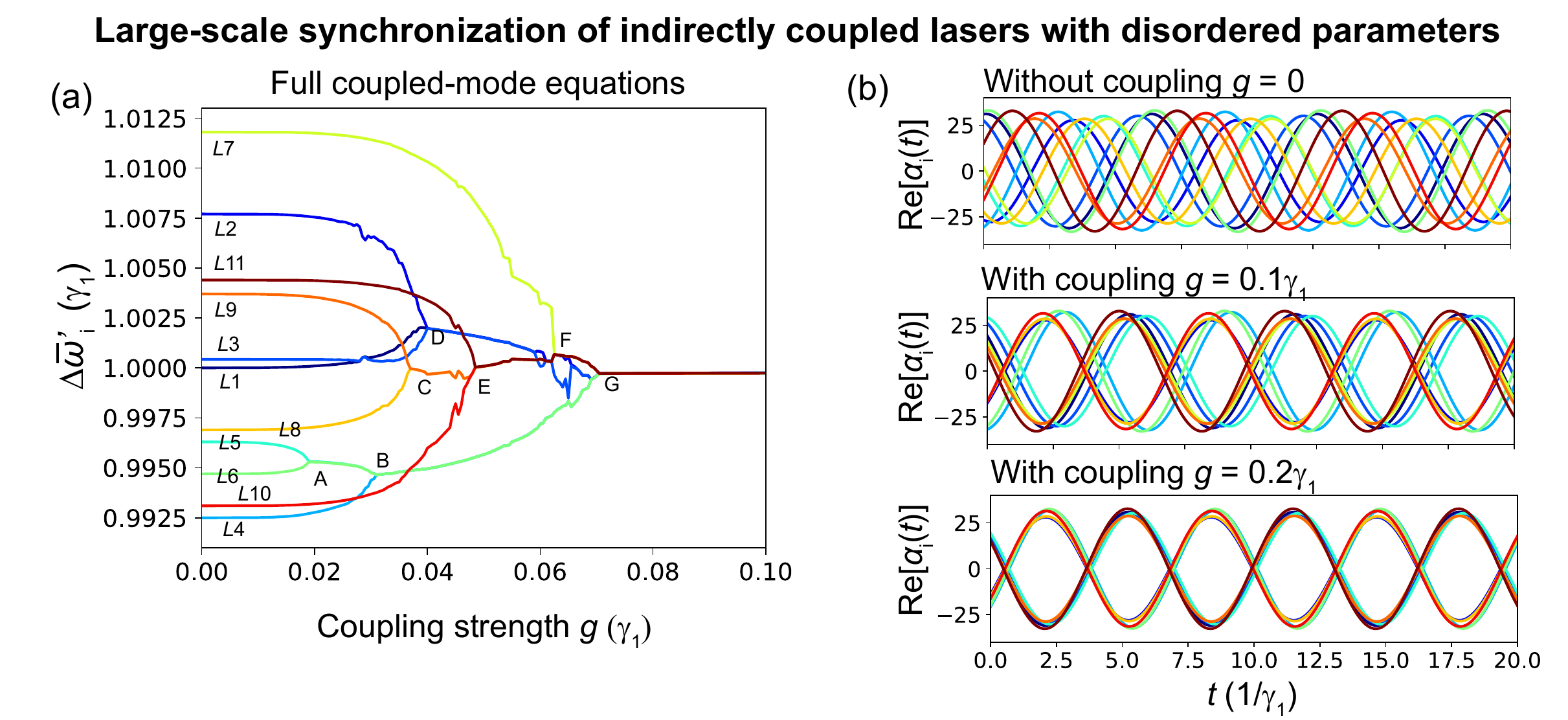}
\caption{Simulations for eleven indirectly coupled lasers with disordered parameters. The parameter values of the laser and cold cavities are summarized in Table \ref{table1}. (a) The mean oscillation frequencies of the eleven lasers $\bar{\omega}_i^\prime$ are shown as a function the coupling strength $g_i=g$ for all $i$. The synchronization points are denoted by A-G in the same way as in Fig. 4(b) in the main text. (b) Time evolutions of the real parts of the fields in all the laser cavities for $g=0$ (top), $0.1\gamma_1$ (middle), and $0.2\gamma_1$ (bottom).}
\label{fig:disorder}
\end{figure}

We simulate eleven indirectly coupled lasers with the same configuration as in Fig. 4(a) in the main text, but all parameters except for coupling strengths $g_i$ have slightly different values. The parameters of all laser ($\epsilon_i$, $\beta_i$, and $\gamma_i$) and cold cavities ($\Omega_i$ and $\Gamma_i$) are randomly distributed around their mean values, which is summarized in Table \ref{table1}. Note that the laser cavities have the same frequencies as those in Fig. 4 in the main text. First, in Fig. \ref{fig:disorder}(a), we show the mean frequencies of the eleven lasers as a function of the coupling strength $g(=g_i)$. The indices of synchronization points A-G are denoted in the same way as in Fig. 4(b) in the main text. The synchronization tree shown in Fig. \ref{fig:disorder}(a) well resembles that in Fig. 4(b) in the main text and clearly indicates that large-scale synchronization can be realized even when the cavities do not have equal parameter values. However, synchronization behavior around synchronization point F is more complicated than that in Fig. 4(b) in the main text. Interestingly, de-synchronization, discussed in Ref. \cite{Zheng1998}, may be observed around point F. Second, in Fig. \ref{fig:disorder}(b), we show the time evolutions of the laser oscillations for $g=0$, $0.1\gamma_1$, and $0.2\gamma_1$ in the top, middle, and bottom panels of Fig. \ref{fig:disorder}(b), respectively. Due to the difference in $\beta_i$ and $\epsilon_i$, Fig. \ref{fig:disorder}(b) indicates that the amplitudes of all the laser oscillations are slightly different. As we expect, synchronization occurs with $g=0.1\gamma_1$ [see the middle panel Fig. \ref{fig:disorder}(b)], but the pair of synchronized oscillations have slightly different phases. When the coupling is increased to $g=0.2\gamma_1$, all the pairs of synchronized oscillations have the same phase [see the bottom panel Fig. \ref{fig:disorder}(b)].

\end{document}